\documentclass[a4paper]{article}
\usepackage[text={6.5in,10in},centering]{geometry}
\usepackage{booktabs}
\usepackage{amsmath,amsfonts,amsthm,amssymb}
\usepackage{algorithmic}
\usepackage{algorithm}
\usepackage{multirow,array}
\usepackage{tabularx}
\usepackage[table, dvipsnames]{xcolor}
\usepackage{caption, subcaption}
\usepackage{textcomp}
\usepackage{stfloats}
\usepackage{url}
\usepackage{verbatim}
\usepackage{hyperref}
\usepackage{cleveref}
\usepackage{graphicx, color}
\usepackage{cite}
\usepackage{authblk}
\crefname{figure}{Fig.}{Figs.}
\Crefname{figure}{Fig.}{Figs.}
\crefname{table}{Table}{Table}
\Crefname{table}{Table}{Table}
    
\newcommand{\eg}{{\it e.g.}}
\newcommand{\ie}{{\it i.e.}}

\newcommand{\Expect}{\mathrm{\bf E{}}}

\newcommand{\reals}{{\mbox{\bf R}}}
\newcommand{\symm}{{\mbox{\bf S}}}
\newcommand{\dom}{\mathop{\bf dom}}
\newcommand{\minimize}{\mathop{\text{minimize}{}}}

\newcommand{\Tr}{\mathop{\bf tr}}
\newcommand{\diag}{\mathop{\bf diag}}
\newcommand{\st}{\text{subject to}}

\newcommand{\argmin}{\mathop{\rm argmin}}

\newcommand{\wm}{\mbox{W/m}^2} 
\newcommand{\kw}{\mbox{kW}}
\newcommand{\kwh}{\mbox{kWh}}
\newcommand{\currency}{\mbox{THB}}

\newcommand{\xc}{x_\text{chg}}
\newcommand{\xdc}{x_\text{dchg}}
\newcommand{\xgi}{x^{(i)}_\text{g}}
\newcommand{\xgh}{x^{(i)}_\text{g,h}}
\newcommand{\xng}{x_\text{ng}}

\newcommand{\xngd}{x_{\text{ng,d}}}
\newcommand{\pg}{P^{(i)}_\text{g}}

\newcommand{\png}{P_\text{ng}}

\newcommand{\pnet}{P_\text{net}}

\newcommand{\pl}{P_\text{load}}
\newcommand{\pr}{P_\text{renew}}

\newcommand{\pc}{P_\text{chg}}
\newcommand{\pcramp}{P_\text{chg,ramp}}

\newcommand{\pdc}{P_\text{dchg}}
\newcommand{\pdcramp}{P_\text{dchg,ramp}}
\newcommand{\soc}{\text{SoC}}

\newcommand{\pgdai}{P^{(i)}_\text{g,da}}
\newcommand{\pgrti}{P^{(i)}_\text{g,rt}}
\newcommand{\pngda}{P_\text{ng,da}}
\newcommand{\pngrt}{P_\text{ng,rt}}
\newcommand{\psda}{P_\text{sell,da}}
\newcommand{\psrt}{P_\text{sell,rt}}

\newcommand{\er}{e_\text{renew}}
\newcommand{\el}{e_\text{load}}
\newcommand{\erbf}{\mathbf{e}_\text{renew}}
\newcommand{\elbf}{\mathbf{e}_\text{load}}

\newcommand{\nsource}{N_\text{source}}
\newcommand{\nscenario}{N_\text{scenario}}
\newcommand{\bg}{b^{(i)}_\text{g}}
\newcommand{\bng}{b_\text{ng}}
\newcommand{\bgda}{b_\text{g,da}}
\newcommand{\bngda}{b_\text{ng,da}}
\newcommand{\bgrt}{b_\text{g,rt}}
\newcommand{\bngrt}{b_\text{ng,rt}}
\newcommand{\sgda}{s_\text{g,da}}

\newcommand{\sgrt}{s_\text{g,rt}}

\graphicspath{{figures/}}

\arrayrulecolor{teal}

\title{Stochastic EMS for Optimal 24/7 Carbon-Free Energy Operations\footnote{This work has been submitted to the IEEE for possible publication. Copyright may be transferred without notice, after which this version may no longer be accessible.}}

\author[2]{Natanon Tongamrak}
\author[3]{Kannapha Amaruchkul}
\author[2]{Wijarn Wangdee}
\author[1]{Jitkomut Songsiri\thanks{Corresponding author.}}

\affil[1]{Department of Electrical Engineering, Faculty of Engineering, Chulalongkorn University, Bangkok, Thailand}
\affil[2]{Center of Excellence in Electrical Power Technology, Faculty of Engineering, Chulalongkorn University, Thailand}
\affil[3]{Graduate School of Applied Statistics, National Institute of Development Administration (NIDA), Bangkok, Thailand}

\affil[ ]{\textit{e-mail: natanon.t@chula.ac.th, kamaruchkul@as.nida.ac.th, wijarn.w@chula.ac.th, jitkomut.s@chula.ac.th}}

\date{}

\begin{document}


\maketitle

\begin{abstract}
This paper proposes a two-stage stochastic optimization formulation to determine optimal operation and procurement plans for achieving a 24/7 carbon-free energy (CFE) compliance at minimized cost. The system in consideration follows primary energy technologies in Thailand including solar power, battery storage, and a diverse portfolio of renewable and carbon-based energy procurement sources. Unlike existing literature focused on long-term planning, this study addresses near real-time operations using a 15-minute resolution. A novel feature of the formulation is the explicit treatment of CFE compliance as a model parameter, enabling flexible targets such as a minimum percentage of hourly matching or a required number of carbon-free days within a multi-day horizon. The mixed-integer linear programming formulation accounts for uncertainties in load and solar generation by integrating deep learning-based forecasting within a receding horizon framework. By optimizing battery profiles and multi-source procurement simultaneously, the proposed system provides a feasible pathway for transitioning to carbon-free operations in emerging energy markets.
\end{abstract}

\begin{flushleft}
\textbf{Keywords:} RE100, 24/7 carbon-free energy, energy management system (EMS), renewable energy, two-stage stochastic programming, mixed integer linear programming
\end{flushleft}

\newpage

\section*{Nomenclature}
\subsection*{Specification}
CF and CB periods denote intervals where consumption is derived exclusively from carbon-free or carbon-based sources, respectively. While nuanced differences exist, \emph{green, renewable}, and \emph{carbon-free} are used interchangeably in this paper. The EMS planning horizon, $D$ (days), consists of $T$ total time steps. Given an operational resolution of 15 minutes, $T = D\times 24 \times \frac{60}{15}$, while the total hourly duration is $H = 24D$.  

\subsection*{Indices}
The optimization variable is indexed by $t=1, 2,\ldots,T$. The Market Time Unit (MTU) is assumed as hourly, and indexed by $h =1, 2,\ldots,H$. Variables associated with days are indexed by $d=1,2,\ldots,D$. There are $N_\text{source}$ green energy sellers; each indexed by $i=1,2,\ldots,N_\text{source}$.

\subsection*{Optimization Variables}
The variables of (deterministic) EMS formulation are:\\[1ex]
\begin{tabular}{ p{0.13\columnwidth} p{0.82\columnwidth}} 
$\xgi,\xng$  &  Binary status of buying green/non-green energy \\
$\xngd$  & Non-green status of day $d$ \\
$u$  & Auxiliary binary variable in selecting CF days \\
$v$ & Auxiliary binary variable in green sourcing rules \\ 
$\pg, \png$  & Purchased green/non-green power ($\kw$) \\
$\xc,\xdc$  & Binary status of charging/discharging battery \\
$\pc,\pdc$ & Charging/Discharging battery power ($\kw$)\\
$\soc$ & Battery's state-of-charge ($\%$) \\
$\pnet$ & Net power at point of common coupling (\kw) \\
$\pnet^{+},\pnet^{-}$ & Positive/negative part of net power (\kw)
\end{tabular} \\[1ex]

All variables except $u, \xngd$ are functions of $t$, while $\soc(t)$ has length $T+1$. The variables $u$ and $\xngd$ are functions of $d$. The variables with superscript $i$ refer to the quantity of the $i^{\mathrm{th}}$ green energy source.

The additional variables for stochastic formulation are grouped into DA (day-ahead stage) and RT (real-time stage):\\[1ex]
\begin{tabular}{ p{0.2\columnwidth} p{0.75\columnwidth}}
$\pgdai$  & Purchased green power in DA ($\kw$) \\
$\pgrti(\cdot ,\xi)$ & Purchased green power in RT ($\kw$) \\
$\pngda$  & Purchased non-green power in DA ($\kw$) \\
$\pngrt(\cdot,\xi)$ & Purchased non-green power in RT ($\kw$) \\
$\psda$ & Selling power in DA markets ($\kw$) \\
$\psrt(\cdot, \xi) $ & Selling power in RT markets ($\kw$) \\
\end{tabular}\\[1ex]

The variables $\pgdai(t), \pngda(t),\psda(t)$ are functions of $t$. The variables $\pgrti(t,\xi), \pngrt(t,\xi), \psrt(t,\xi)$ are functions of $t$ and scenario parameter $\xi \in \mathcal{S}$ where $\mathcal{S}$ is the scenario set with cardinality of $\nscenario$. The variables with superscript $i$ refer to the quantity of the $i^{\mathrm{th}}$ RE source.

\subsection*{Problem parameters}
\begin{flushleft}
\begin{tabular}{ll}
$\bg,\bng$ & Import rate of green/non-green energy ($\currency/\kwh$)\\
$s$ & Export rate of energy ($\currency/\kwh$) \\
$\pr,\pl $ & Renewable energy/Load (\kw)\\
$\eta_c, \eta_d$ & Charging/discharging efficiency \\
$\pcramp$ & Charging ramp rate limit (\kw)\\
$\pdcramp$ & Discharging ramp rate limit (\kw)\\
$M$ & a large constant \\
\end{tabular}
\end{flushleft}
The parameters $\bg(t), \bng(t), s(t), \pr(t), \pl(t)$ are functions of $t$. Other parameters are scalars. The parameter $\bg$ is the green energy price from the source $i$.

\section{Introduction}
The concept of 100\% renewable energy (RE) was established to drive global decarbonization by matching \emph{annual} electricity demand with equivalent renewable generation. As the global transition progresses, solar and wind have emerged as the primary technological pillars. To address the inherent variability of these sources, future research can focus on diverse storage solutions, power-to-X technologies, and the integration of carbon dioxide removal (CDR) to achieve net-negative emissions \cite{Breyer2022}. To reach 100\% RE targets, Renewable Energy Certificates (RECs) were introduced to track clean generation. While RECs treat energy as a flexible commodity, they often mask hourly supply-demand misalignments. Thus, a 100\% RE goal fails to guarantee a zero-carbon grid by ignoring the spatial and temporal realities of electricity production versus consumption \cite{Miller2020, Peninsula2023}.

Moving beyond annual 100\% RE annual volume matching, Google proposed the 24/7 Carbon-Free Energy (CFE) framework to match hourly electricity consumption with carbon-free generation \cite{Google2472018}. The company aims to operate all data centers on this basis by 2030 \cite{Google247vision2020, Google247roadmap2022}. This roadmap focuses on diversifying renewable sources, optimizing existing production, scaling battery storage, deploying ML algorithms to predict RE, and active demand management. The shift from 100\% RE towards the more challenging 24/7 CFE would promote the development of clean technologies, storage and transmission to create a more robust framework for a full decarbonization grid \cite{Aagaard2023}. The IRENA Coalition for Action defines 100\% RE as the total exclusion of nuclear and fossil fuels, requiring 24/7 hourly renewable matching by 2050 \cite{IRENA2024}. Achieving this granularity requires fossil fuel phase-outs, accelerating the electrification of end-use sectors, and enhanced grid flexibility and efficiency.

A substantial body of work has examined the feasibility and cost of 100\% renewable or 24/7-style electricity systems using long-horizon, hourly energy balance models and capacity optimization \cite{Cheng2022renewJapan, Riepin2024}. These planning-oriented studies provide insights into optimal capacity mixes, storage requirements, and system-wide decarbonisation benefits, but primarily address system configuration or energy technologies rather than real-time operational decision-making under uncertainty. For example, the work of \cite{Bastos2023} considered diesel as a backup for a system with varying renewable and battery capacities. Assuming perfect foresight over a one-year horizon, the study found that a minimum configuration of 400 kW PV and 2 MWh BESS eliminated the need for diesel, achieving 100\% renewable operation. Recent work on hourly renewable power purchase agreements (PPAs) and 24/7 procurement shows that enforcing continuous supply substantially alters capacity mix and costs; however, the proposed formulation remains focused on deterministic contract design rather than near real-time operational control \cite{Jain2022}. The analysis relied on historical renewable energy profiles obtained from meteorological databases, rather than forecasting models. Overall, previous planning-focused analyses do not address how 24/7 CFE targets can be achieved operationally under short-term uncertainty in load and renewable generation.

Current 24/7 CFE research largely focuses on long-term planning, often overlooking the operational challenges of short-term uncertainty. Some operational study relies on deterministic or rule-based dispatch strategies to minimize cost, without explicitly addressing uncertainty or time-resolved clean energy matching \cite{Al2021}. Recent work has formulated 24/7 procurement as a probabilistic portfolio optimization problem that enforces hourly matching at an annual decision scale, without energy modeling and considering near real-time operational adaptation \cite{Ludkovski2024}. Learning-based forecasting and uncertainty-aware optimization have emerged for specific flexible loads like data centers \cite{Yang2025} where CFE compliance relies on chance constraints. For operational planning, stochastic and receding-horizon control frameworks are often applied for real-time tracking. The work in \cite{Qureshi2021} applied two-stage optimization for tracking pre-declared wind-power dispatch plans where the error of generation-load matching is controlled via an infinity-norm constraint. CVaR-based risk measures of power imbalance was proposed in \cite{Chen2024} for hydrogen-based flexible loads. While effective for general grid operation, these approaches lack explicit mechanisms for time-resolved 24/7 CFE compliance tracking. 

Achieving 24/7 CFE is highly contingent upon specific system configurations and technology portfolios. Current literature suggests three primary pathways to viability: the scaling of renewable capacity and long-duration storage, the application of robust optimization to manage RE source intermittency, and the integration of demand-side management for load flexibility. In Thailand’s energy landscape, 24/7 CFE viability primarily depends on a technology mix of solar power, battery storage, and strategic renewable energy procurement. However, achieving strict hourly matching is often hindered by prohibitive capital costs associated with over-scaling these assets. This paper evaluates a more economically feasible pathway for Thailand by assessing the trade-off between explicit CFE compliance levels and total operational costs within two-stage stochastic optimization and a receding-horizon framework. Operating at a near real-time resolution of 15 minutes, we define compliance as a fraction of carbon-free ‘slots’ over a multi-day horizon. By quantifying compliance as a percentage of total hours (e.g., 90\% hourly matching) or as a target number of CF days (e.g., five CF days within a seven-day horizon), our energy management system (EMS) wisely selects the optimal subset of CF compliant slots to minimize costs. Through this approach, we determine the ideal battery profiles and procurement plans required to meet specific targets. When the compliance degree is set to 100\%, the optimal operational plan yields a 24/7 CFE goal. 

The main contributions of this paper are the following.
\begin{enumerate}
\item \textbf{Strategic compliance formulation:} We introduce a two-stage stochastic optimization model that identifies the most cost-effective `carbon-free interval’ within a given horizon. Unlike rigid matching frameworks, this approach optimizes the selection of compliant intervals to balance operational costs with CFE targets.
\item \textbf{Multi-scale procurement portfolio framework:} We formulate procurement constraints that accommodate diverse budgetary limits and resource availability across different generation sources. Crucially, the model decouples the temporal resolution of procurement updates from the 15-minute operational dispatch, allowing for a realistic simulation of how periodic contractual commitments interact with near real-time EMS.
\item \textbf{Real-time predictive control:} We integrate deep-learning and machine learning-based forecasting with a receding-horizon control (RHC) framework. Operating at a 15-minute resolution, this system accounts for the stochastic nature of solar irradiance and demand, providing a robust tool for analyzing the trade-off between CFE compliance and total system expenditure.
\end{enumerate}

\section{Optimization Formulation}
\textbf{Definition:} A system operates in a carbon-free (CF) setting during an interval if its demand is satisfied exclusively via self-sufficiency or green energy procurement. Given a 15-minute operational resolution, a CF day and a CF hour require this condition to be met for all 96 and 4 constituent intervals, respectively.

\textbf{Definition:} The daily CF compliance level is defined as the requirement to maintain CF operations for $X$ days within a $Y$-day period. Similarly, the hourly CF compliance level specifies a requirement of $X$ CF hours over a $Y$-hour horizon.

As the formulations for daily and hourly CF compliance are structurally similar, the following section presents the daily compliance constraint, as it provides a more intuitive benchmark, while remaining generalizable to hourly CF. We address the optimization of an energy management system consisting of a battery storage unit, self-generated renewable energy sources (\eg, wind, solar), and an electrical load. The system maintains a grid connection, offering access to both green and non-green energy resources and can export excess power for making revenue. The objective is to develop a control mechanism that minimizes the net procurement cost of grid-supplied energy over a $D$-day planning horizon. Specifically, the mechanism will determine i) optimal battery charge/discharge cycles, ii) the requisite quantities of green and non-green energy to be drawn from the grid, and iii) the amount of excess energy to be exported. A key constraint is \textbf{the achievement of the daily CF compliance level}. The solution will also provide iv) the identification of the subset of $D$ days that satisfy a targeted daily CF level.

The problem formulation for the no-export policy where excess energy cannot be sold back to the grid, and its corresponding numerical results are provided in the Supplementary Material.

\subsection{Decision variable}
Assume that there are a number of CF resources, set as $\nsource$, that offer green energy at different prices. Denote $\xgi(t)$ and $\xng(t)$ the buying green from source $i$ and non-green energy status (binary value) at time $t$, respectively. Denote $\pg(t)$ and $\png(t)$ the purchased green power from source $i$ and non-green power at time $t$, respectively.

When operation granularity is 15 minutes, but the Market Time Unit is on hourly basis, so the quantity of purchased power at each $t$ must be equal within an hour. Using $h$ as an hour index, this creates constraints for each $h=1,2,\ldots,H$,
\begin{equation}
z(t) \;\;\text{for all $t \in \{ 4(h-1)+1:4h \}$ are equal}
\label{eq:hourly_var}
\end{equation}
where the condition \eqref{eq:hourly_var} applies to $\xng, \pngda, \psda$, and to $\xgi(t),\pgdai$ for $i=1,\ldots,\nsource$.

Sub-day variable: Denote $\xngd(d)$ a sub-block of $\xng(t)$ whose time indices lie within day $d$. For example, the group of day 1 is $\xngd(1) = (\xng(1),\ldots,\xng(96))$ and the group of day 2 is $\xngd(2) = (\xng(97),\ldots,\xng(192))$. For each $d =1,2,\ldots,D$, $\xngd$ and $\xng$ are linked by
\begin{equation}
\xngd(d) = \xng(96(d-1)+1:96d).
\label{eq:dayblock}
\end{equation}

\subsection{Objective}
\label{sec:obj}
The objective is to minimize the total net cost, calculated as the total expense for purchased power minus the revenue generated from power sold back to the grid.
\[
\text{Net cost} = \text{Purchasing cost} - \text{Selling revenue}.
\]
Let $\bg(t),\bng(t)$ be the unit purchase price of green/non-green energy, respectively at time $t$, and let $s(t)$ be the export rate from selling surplus energy back to the grid. The purchase price for green energy is assumed to be higher than that of non-green energy, $\bg(t) > \bng(t)$ for all $i$. Furthermore, the rate for energy exported (sold) to the grid is lower than both energy import prices, $s(t) < \bng(t), \bg(t)$, for all $i$.

Let $\pnet(t)$ be the net power exchanged at the system's point of common coupling (PCC). To model the simultaneous inability to both buy and sell power at a given time $t$, we introduce two auxiliary variables: $\pnet^{+}(t)$, the positive part representing the total power purchased from sources, and $\pnet^{-}$, the negative part being the total power exported to the grid. The objective as net cost is
\begin{equation}
\text{Net cost} = \Delta t \sum_{t=1}^T  \left (  \bng(t) \png(t) +  \sum_{i=1}^{\nsource} \bg(t) \pg(t) \right )
 - \Delta t\sum_{t=1}^T  s(t) \pnet^{-}(t).
 \label{eq:netcost}
\end{equation}
where $\png(t) + \sum_{i=1}^{\nsource} \pg(t)  = \pnet^{+}$. We must force the two auxiliary variables to accurately represent the positive/negative parts of $\pnet(t)$ using the following constraints:
\begin{gather*}
\pnet(t) = \pnet^{+}(t)-\pnet^{-}(t), \quad \pnet^{+}, \pnet^{-} \geq 0,\quad |\pnet(t)| = \pnet^{+}+\pnet^{-}.
\end{gather*}
However, since $| \cdot |$ introduces a nonlinearity, the second equality is typically relaxed to an inequality: 
\begin{equation}
|\pnet(t)| \leq \pnet^{+}+\pnet^{-},
\label{eq:abspnet}
\end{equation}
which can be further cast as linear inequalities. Due to the nature of objective function where the buy rate is always higher than the sell rate, the optimization naturally drives the inequality~\eqref{eq:abspnet} to become tight at optimum. This ensures that at any given time, exactly one $\pnet^{+}(t)$ or $\pnet^{-}(t)$ will be non-zero.

\subsection{Constraints}
The list of constraints is as follows.
\paragraph{Battery model.} The charging and discharging power must follow the battery status. 
\begin{eqnarray}
\pc(t) &\leq& \text{Charge rate}_{\max} \cdot \xc(t) , \label{eq:chargemax} \\
\pdc(t) &\leq& \text{Discharge rate}_{\max} \cdot \xdc(t), \label{eq:dischargemax}
\end{eqnarray}
to prevent sudden change of charging/discharging power, for $2,3,\ldots,T$, the difference of power between two consecutive time steps is limited. For each $t=1,2,\ldots,T$,
\begin{eqnarray}
-\pcramp &\leq& \pc(t) - \pc(t-1) \leq \pcramp, \label{eq:charge_slew}\\
- \pdcramp &\leq& \pdc(t) - \pdc(t-1) \leq \pdcramp. \label{eq:discharge_slew}
\end{eqnarray}
Battery status must be either charged or discharged. 
\begin{equation}
0 \leq \xc(t) + \xdc(t) \leq 1, \quad t=1,2,\ldots,T.
\label{eq:chg_dchg}
\end{equation}
When the battery is charged or discharged, this updates the state-of-charge (SoC) dynamic of the battery as follows. For $t=1,2,\ldots,T$
\begin{equation}
\soc(t+1) = \soc(t)
+ \frac{100}{\text{BattCapacity}} \left ( \eta_c \pc(t) \Delta t - \frac{\pdc(t) \Delta t}{\eta_d} \right ).
\label{eq:battsoc}
\end{equation}
The SoC must stay within a range to preserve its life time.
\begin{equation}
\soc_{\min} \leq \soc(t) \leq \soc_{\max},\quad t=1,2,\ldots,T.
\label{eq:battminmax}
\end{equation}
In order to maintain the battery's SoC for the next operation day, we can set the SoC at the end of each day to a desired value. Let $\mathcal{I}_{\text{last hour}}$ be the index set of the last hour of each day. The constraint is simply linear as 
\begin{equation}
\soc(k) \geq \text{Terminal SoC}, \quad k \in \mathcal{I}_{\text{last hour}}
\label{eq:terminal_soc}
\end{equation}

\paragraph{Power balance.} Total power generation is the renewable power $\pr(t)$ and the discharging power from the battery, $\pdc(t)$, while the total demand is the electrical load $\pl$ and the charging power from the battery $\pc$. 
\begin{equation}
\pnet(t) = \pl(t) + \pc(t) - \pr(t) - \pdc(t),
\label{eq:powerbalance_deterministic}
\end{equation}
for $t=1,2,\ldots,T$. Note that $\pr$ and $\pl$ in \eqref{eq:powerbalance_deterministic} are typically provided from forecast models (predictions in the future). The variable $\pnet$ denotes the net power as total load subtracted by generation. If $\pnet(t) > 0$, the system draws additional power from the grid; on the other hand, if $\pnet(t) < 0$, the system can sell the energy back to the grid. 

\paragraph{Imported and exported power.} The imported power is supplied by purchasing from green and/or non-green power sources, which introduces the following constraint:
\begin{equation}
\pnet^+(t) = \png(t) + \sum_{i=1}^{\nsource} \pg(t) ,\;\; t=1,\ldots,T.
\label{eq:buypower1}
\end{equation}
The modeling part explained in \Cref{sec:obj} additionally enforces the following constraints for $t=1,\ldots,T$,
\begin{gather}
\pnet(t) = \pnet^{+}(t) - \pnet^{-}(t), \;\;\pnet^{+}, \pnet^{-} \geq 0,  \label{eq:pnet1} \\
-(\pnet^{+}(t) + \pnet^{-}(t)) \leq \pnet(t) \leq \pnet^{+}(t) + \pnet^{-}(t). \label{eq:pnet2}
\end{gather}

\paragraph{Green power sourcing rules.} 
The quantity of green power purchased must be $0$ if the green status variable is $0$, and must be non-negative (up to $M$) if the green status is $1$. The same relationship applies to the non-green power and its status variable. For $t=1,2,\ldots,T$,
\begin{equation}
\begin{split}
\epsilon \xgi(t) \leq \pg(t) &\leq M \xgi(t),\;\; i=1,2,\ldots, \nsource,\\
\epsilon \xng(t) \leq \png(t) &\leq M \xng(t), \;\; \text{$\epsilon$ is a small number}.
\end{split}
\label{eq:green_sourcing1}
\end{equation}
Moreover, we design the green and non-green power sources to be mutually exclusive; they cannot both be purchased simultaneously. Since we can buy green energy from more than one source at a time, we can enforce
\[
0 \leq \xng(t) + \max_{i=1,\ldots,\nsource} \xgi(t)\;\; \leq 1,\;\; t=1,2,\ldots,T,
\]
which can be cast as linear inequalities using a variable $v(t)$:
\begin{gather}
\xgi(t) \leq v(t), \;\; i=1,2,\ldots,\nsource, \;\;t=1,2,\ldots,T, \label{eq:green_sourcing2} \\
0 \leq v(t) + \xng(t) \leq 1, \;\; t=1,2,\ldots,T.
\label{eq:green_sourcing3}
\end{gather}

\paragraph{Green energy budget constraint.} Green energy from each source is limited by an assigned allocation of energy (\kwh).
\begin{equation}
\sum_{t=1}^T \pg(t) \Delta t \leq \text{allocation}_i, \quad i=1,2,\ldots,\nsource
\label{eq:greenbudget}
\end{equation}

\paragraph{Selecting CF days.}
For a planning horizon of $D$ days with a required CF compliance level, we introduce a binary selection variable $u \in \{0,1\}^D$ indexed by day $d$. Let $u_d = 0$ denote a CF day, where the system is restricted to self-supplied battery or purchased green power; otherwise, $u_d = 1$ denotes a carbon-based (CB) day. 

CB day count: The total number of CB days must not exceed the maximum allowed CB days.
\begin{equation}
\sum_{d=1}^D u_d \leq \text{no. of CB days}. 
\label{eq:nonREdays}
\end{equation}
Linking constraint: The variable $u_d$ serves as an upper bound for the non-green status variable on day $d$. This means that non-green power can be positive only when $u_d = 1$.

\begin{equation}
\Vert \xngd(d) \Vert_\infty \leq u_d,\; d=1,2,\ldots,D.
\label{eq:max_nongreen}
\end{equation}
The inequality on $\Vert \cdot \Vert_\infty$ can be typically cast as linear inequalities from the $\max(\cdot)$ operation.

\subsection{Flexible-CFE optimization formulation}
The F-CFE (Flexible Carbon-Free Energy) formulation optimizes the selection of CF days within a $D$-day horizon based on daily compliance requirements, while determining the optimal procurement and battery operation strategy. 
\begin{equation}
\begin{array}{ll}
\minimize & \text{Net cost} \;\;\eqref{eq:netcost} \\
\st & \text{Power balance}\;\; \eqref{eq:powerbalance_deterministic} \\
& \text{Exported/Imported power relation} \;\; \eqref{eq:buypower1},\eqref{eq:pnet1},\eqref{eq:pnet2}, \\
& \text{Green sourcing rule}\;\; \eqref{eq:green_sourcing1},\eqref{eq:green_sourcing2}, \eqref{eq:green_sourcing3} \\
& \text{Green energy budget} \;\; \eqref{eq:greenbudget} \\
& \text{Hourly and day-block relation}\;\; \eqref{eq:hourly_var},\eqref{eq:dayblock} \\
& \text{Battery constraints}~\eqref{eq:chargemax}-\eqref{eq:terminal_soc}, \\
& \text{Selecting CF days}, \eqref{eq:nonREdays},\eqref{eq:max_nongreen}
\end{array}
\label{eq:cfeopt}
\end{equation}
The problem~\eqref{eq:cfeopt} is a mixed-integer linear program (MILP) due to the binary variables governing the status of green/non-green power and the battery's charging/discharging modes. The introduction of the green energy budget's constraint~\eqref{eq:greenbudget} may lead the problem to infeasibility if the specified budget is insufficient to meet a high targeted number of CF days. Removing the budget constraint~\eqref{eq:greenbudget} ensures problem feasibility by implying that sufficient green resources are always available, thereby guaranteeing the achievement of the targeted CF days with minimum cost. To solve~\eqref{eq:cfeopt} numerically, we employ \texttt{cvxpy} with Gurobi solver \cite{cvxpy}.

\subsection{Rolling F-CFE EMS}
Suppose we aim to implement the F-CFE EMS on a long period of interest, \eg, months or a year. The rolling EMS simulation uses the model predictive control (MPC) principle and problem~\eqref{eq:cfeopt} to stride on a daily basis. Starting on day $d=1$, the rolling EMS involves three main steps.

\paragraph{Prediction:} Forecast the $D$-day-ahead $\pr, \pl$ required for the power balance constraints~\eqref{eq:powerbalance_deterministic}.
\paragraph{Optimization:} Solve~\eqref{eq:cfeopt} to get the optimal $D$-day-ahead plan, including CF day selection, power purchasing, power selling, and battery operations. 
\paragraph{Realize and Update:} Apply the action plan for the \emph{first day} only. Update the system status using real-time measurements: i) $\pnet$: Updated in~\eqref{eq:powerbalance_deterministic} using  actual $\pr,\pl$, ii) $\pg,\png$: Adjusted based on the committed CF/CB status ($\xgi,\xng$) and the adjusted amount required by~\eqref{eq:buypower1}.
 
Set $d:= d+1$ and repeat until the end of study period. This daily measurement update and re-planning provides the optimal rolling solution for CF planning. 

\section{Forecasting models}
\label{sec:forecast}
To perform any planning tasks, forecasting models are important for anticipating trends in both generation and demand sides. We explore machine learning based models, which are \textbf{LightGBM}, \textbf{LSTM} and \textbf{Neural Prophet} to generate a deterministic forecast of $\pl$ and $\pr$. The LSTM model architecture follows Figure 11 in \cite{piestECMX2025} but trained with the MAE loss.

Both solar power and load forecasting models rely on a common model setting that the input and output $(y)$ are related by $y(t+k|t) = f(y(t-L_y:t),x(t-L_x:t),z(t+k|t))$ where $y(t-L_y:t)$ represents lagged values of $y$ from time $t-L_y$ up to time $t$; $x(t)$ are lagged exogenous variables measured from time $t-L_x$ up to time $t$; and $z(t+k|t)$ are future regressors for time $t+k$, available at time $t$.

\subsection{Model features}
\paragraph{Electrical load forecasting model.} The regressors are historical load, and future regressors from weather forecasts \cite{sodapro}, and building usage calendar. 

\paragraph{Solar power forecasting model.} The solar power forecasting follows an indirect approach: solar irradiance $I$ is first predicted and then converted to power output via: $\alpha$: $\pr(t) = \alpha I(t)$ where $\alpha$ is an irradiance-to-power conversion factor estimated from historical data. The irradiance forecasting model requires historical irradiance, and future regressors from weather forecasts~\cite{sodapro}, and clear-sky irradiance (Ineichen model with parameters adjusted to Thailand.) \cite{solarmap2025}

Model hyperparameters, such as the number of historical lags and selected calendar features, are tuned through model selection and hyperparameter optimization.

\subsection{Forecasting results}
Load consumption and solar irradiance forecasts are developed using 15-minute resolution data
spanning from March 2023 to May 2024. The models are evaluated on the period from
June 2024 to May 2025. A 7-day-ahead forecast horizon is employed with $T=672$ time steps. The models are evaluated with daily rolling forecasts.

\begin{figure}
    \centering
    \includegraphics[width=0.5\linewidth]{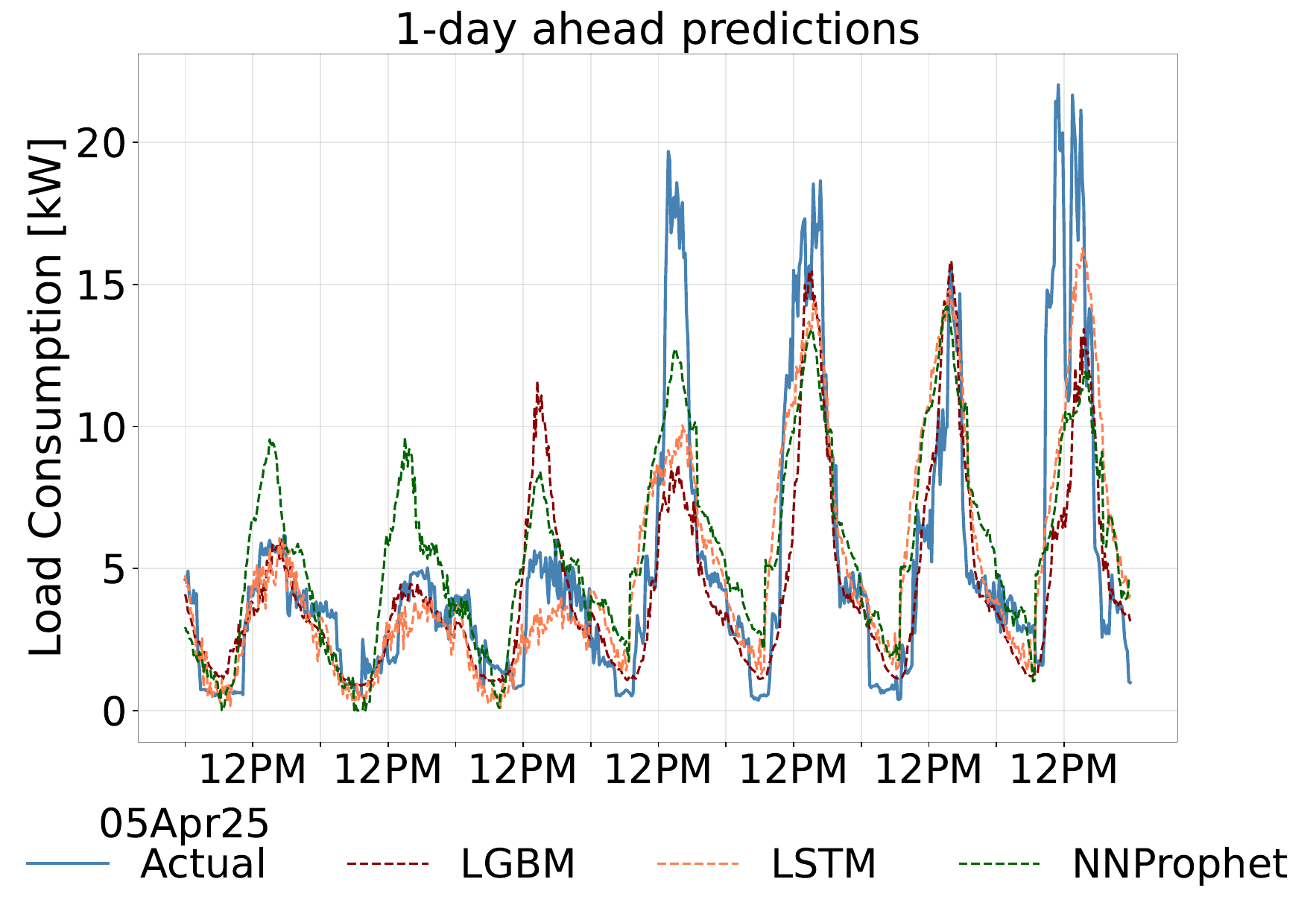}
    \caption{1 day-ahead \textbf{load} forecasts.}
    \label{fig:pl_da_models_ts}
\end{figure}

\paragraph{Load forecasting.} The example of 1 day-ahead load forecast from models choices is shown in \cref{fig:pl_da_models_ts}. and the overall 7 day-ahead forecasting performance is summarized in \cref{tab:pl_da_models_scores}.

\begin{table}
\centering
\caption{Overall 7-day \textbf{load} forecasting metrics.}
\label{tab:pl_da_models_scores}
\begin{tabular}{lrrr}
\toprule
 & LGBM & LSTM & NNProphet \\
\midrule
MAE (\kw) & \textbf{1.72} & 1.75 & 1.85 \\
NMAE (\%) & \textbf{4.92} & 5.02 & 5.29 \\
RMSE (\kw) & 2.68 & \textbf{2.61} & 2.61 \\
NRMSE (\%) & 7.68 & \textbf{7.48} & 7.49 \\
MBE (\kw) & -0.44 & -0.47 & \textbf{-0.43} \\
NMBE (\%) & -1.27 & -1.34 & \textbf{-1.23} \\
sMAPE (\%)& \textbf{6.88} & 7.70 & 9.95 \\
\bottomrule
\end{tabular}
\end{table}

\begin{figure}
    \centering
    \includegraphics[width=0.5\linewidth]{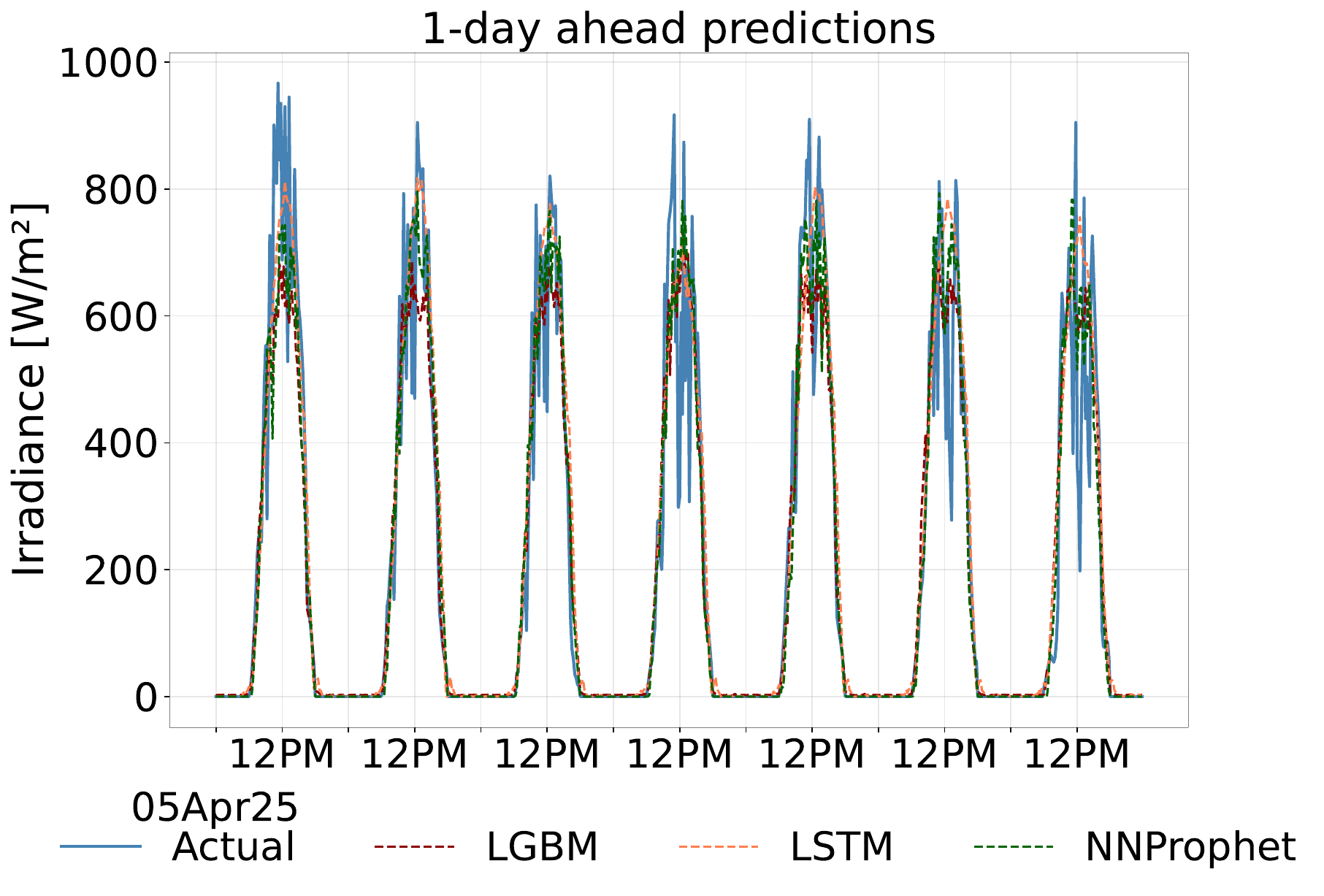}
    \caption{1 day-ahead \textbf{solar irradiance} forecasts.}
    \label{fig:ir_da_models_ts}
\end{figure}

Although all models achieve close performance, LightGBM demonstrates the best overall accuracy, achieving the lowest MAE, NMAE, and sMAPE. As for RMSE metric, the LSTM slightly outperforms LightGBM. 
\paragraph{Irradiance forecasting}
An example of 1 day-ahead solar irradiance forecasts is shown in \cref{fig:ir_da_models_ts} and the overall 7 day-ahead forecasting performance is summarized in \cref{tab:ir_da_models_scores}.

\begin{table}
\centering
\caption{Overall 7-day \textbf{irradiance} forecasting metrics.}
\label{tab:ir_da_models_scores}
\begin{tabular}{lrrr}
\toprule
 & LGBM & LSTM & NNProphet \\
\midrule
MAE ($\wm$) & \textbf{99.32} & 100.99 & 101.02 \\
NMAE (\%) & \textbf{26.29} & 26.73 & 26.74 \\
RMSE ($\wm$) & \textbf{142.75} & 147.05 & 142.98 \\
NRMSE (\%) & \textbf{37.78} & 38.92 & 37.84 \\
MBE ($\wm$) & \textbf{-7.78} & 24.78 & -8.82 \\
NMBE (\%) & \textbf{-2.06} & 6.56 & -2.34 \\
sMAPE (\%) & \textbf{5.03} & 5.42 & 7.00 \\
\bottomrule
\end{tabular}
\end{table}

For solar irradiance forecasting, LightGBM achieves the best results in all metrics, while LSTM shows higher errors in MBE and RMSE. For this reason, we chose LightGBM forecasts for subsequent results.

\section{Stochastic Flexible CFE formulation}
The two-stage stochastic formulation is designed to commit first-stage operations on the day-ahead (DA) horizon, minimizing the net cost objective. This is done while explicitly accounting for the uncertainties in the forecasted values of $\pr$ and $\pl$, which are modeled via a scenario set $\mathcal{S}$. The objective function is the sum of the first-stage cost and the expected cost of the second-stage (real-time, RT) adjustments. These second-stage decisions are made once the true realization of the uncertainty is revealed \cite{BirgeLouveaux2011,alharbi2017}.

\subsection{Variables and objective}
The optimization variables for the stochastic formulation are categorized into the first-stage (or committed) and the second-stage (or recourse) variables: $x_{\text{1st}}= (\xgi, \xng, \pgdai, \pngda, \psda, \pc, \pdc,\xgh,\xngd, u,v)$ and $x_{\text{2nd}}(\xi) = (\pgrti ,\pngrt ,\psrt )$. We have introduced $\psda$ and $\psrt$ as the exported power in DA and RT, respectively.

\paragraph{First-stage (day-ahead) cost.} The first-stage decision, $ x_{\text{1st}}$, represents the committed solution made in the DA market before any uncertainty is revealed. The net cost is given by
\[
f_{\text{da}}(x_{\text{1st}}) = \Delta t \sum_{t=1}^T \left [ \bngda(t) \pngda(t) + \sum_{i=1}^{\nsource} \bgda(t) \pgdai(t) \right ] - \Delta t \sum_{t=1}^T \sgda(t) \psda(t).
\]
\paragraph{Second-stage (real-time) cost.} The second-stage decision, $x_{\text{2nd}}$, represents the operational adjustments made in RT for a given realization of uncertainty, $\xi \in \mathcal{S}$. These adjustments minimize the cost incurred after the true values of uncertain parameters (renewable and load) are revealed. The net cost for a fixed scenario $\xi$ is
\begin{equation}
f_{\text{rt}}(x_{\text{2nd}}(\xi)) = \Delta t\sum_{t=1}^T \left [  \bngrt(t) \pngrt(t,\xi) +  \sum_{i=1}^{\nsource} \bgrt(t) \pgrti (t,\xi) \right ] - \Delta t\sum_{t=1}^T   \sgrt(t)\psrt(t,\xi) .
\end{equation}
The overall objective of the stochastic program then minimizes the first-stage cost plus the expected value of the second-stage cost across all scenarios.
\begin{equation}
\text{objective} = f_{\text{da}}(x_{\text{1st}}) + \Expect_{\xi}[ f_{\text{rt}}(x_{\text{2nd}}(\xi)) ].
\end{equation}
\subsection{Constraints}
In the stochastic program, the deterministic constraints of problem~\eqref{eq:cfeopt} are partitioned and adjusted to maintain scenario-wise feasibility. All operational and physical constraints involving the second-stage variable, $x_{\text{2nd}}(\xi)$, must be satisfied for all scenarios $\xi \in \mathcal{S}$. Non-anticipativity constraints enforce that first-stage decision variables are identical across scenarios.
 
\paragraph{Stochastic power balance.} 
The supply (imported power, battery's discharging power, and renewable) must meet the demand (exported power, and load) for all realizations of renewable and load forecasts. For $t=1,\ldots,T$ and for all $\xi \in \mathcal{S}$, 
\begin{multline}
    \sum_{i=1}^{\nsource} [ \pgdai(t) + \pgrti(t,\xi)] + \pngda(t) + \pngrt(t,\xi) 
    + \pdc(t) + \pr(t,\xi) \\= \pc(t) + \psda(t) + \psrt(t,\xi) + \pl(t,\xi).
\label{eq:powerbalance_stoch}
\end{multline}
There are $T \cdot \nscenario$ constraints.

\paragraph{Stochastic green power sourcing rule.} As we let the status of either buying green or non-green power to be the first-stage variable, any adjustment of real-time purchased power must follow the committed purchasing status. For all $\xi \in \mathcal{S}$,
\begin{equation}
\begin{split}
\epsilon \xgi(t) \leq \pgdai(t), \pgrti(t,\xi) &\leq M \xgi(t),\;\; i=1,2,\ldots, \nsource,\\
\epsilon \xng(t) \leq \pngda(t), \pngrt(t,\xi) &\leq M \xng(t).
\end{split}
\label{eq:green_sourcing1_stoch}
\end{equation}
There are $T \cdot \nscenario (\nsource  + 1) $ constraints.
\paragraph{Stochastic green energy budget constraint.} Green energy from each source is limited by an assigned allocation of energy (\kwh). For $i=1,2,\ldots,\nsource$, and for all $\xi \in \mathcal{S}$,
\begin{equation}
\Delta t  \sum_{t=1}^T \pgdai(t) + \pgrti(t,\xi) \leq \text{allocation}_i.
\label{eq:greenbudget_stoch}
\end{equation}
There are $\nsource \cdot \nscenario$ constraints.

\subsection{Stochastic Flexible CFE formulation}
The stochastic F-CFE program can be presented as
\begin{equation}
    \begin{array}{ll}
    \minimize &  f_{\text{da}}(x_{\text{1st}}) + \Expect_{\xi} [ f_{\text{rt}}(x_{\text{2nd}}(\xi)) ]   \\
    \st & \text{Stochastic power balance}\;\; \eqref{eq:powerbalance_stoch}, \\
    & \text{Stochastic green power sourcing}\;\; \eqref{eq:green_sourcing1_stoch}, \eqref{eq:green_sourcing2}, \eqref{eq:green_sourcing3},\\
    & \text{Stochastic green energy budget} \;\; \eqref{eq:greenbudget_stoch}, \\
& \text{Hourly and day-block relation}\;\; \eqref{eq:hourly_var}, \eqref{eq:dayblock} \\
& \text{Battery constraints}~\eqref{eq:chargemax}-\eqref{eq:terminal_soc}, \\
& \text{Selecting CF days}\;\; \eqref{eq:nonREdays},\eqref{eq:max_nongreen}.
    \end{array}
    \label{eq:cfeopt_stoch}
\end{equation}
The stochastic F-CFE approach offers an advantage over deterministic planning by explicitly incorporating the uncertainties inherent in the renewable generation and load. By minimizing the expected net cost averaged cost across scenarios, the formulation yields a first-stage solution that is more robust to future uncertainties. This ensures the committed operations are optimal in expectation, providing a superior action against potential high-cost events that might arise in a single scenario. 

Meanwhile, solving~\eqref{eq:cfeopt_stoch} is numerically challenging. Scenario-wise feasibility requirements make the stochastic F-CFE more prone to infeasibility than \eqref{eq:cfeopt}. The stochastic green energy budget constraint~\eqref{eq:greenbudget_stoch} is the critical factor; if any scenario $\xi$ combines low $\pr$ with high load $\pr$, the necessary commitment of green energy may exceed the budget, leading to an infeasible second stage. The inclusion of $\nscenario$ scenarios scales the problem size directly. Specifically, the dimensions of the second-stage variables ($\pgrti,\pngrt,\psrt$) and the number of constraints~\eqref{eq:powerbalance_stoch}, \eqref{eq:green_sourcing1_stoch}, \eqref{eq:greenbudget_stoch} are all multiplied by $\nscenario$. This parameter is selected to balance computational complexity with significant net-cost improvements.

\subsection{Scenario generation}
Realizations of $\pr(t,\xi)$ and $\pl(t,\xi)$ are assumed to be perturbations from the point forecasts by a random variable.
\begin{subequations}\label{eq:scenario_gen}
    \begin{align}
    \pr(t,\xi) &= \hat{P}_\text{renew}(t) + \er (t, \xi), \label{eq:pr_scenario_gen} \\
    \pl(t,\xi) &= \hat{P}_\text{load}(t) + \el(t,\xi). \label{eq:pl_scenario_gen}
    \end{align}
    \end{subequations}
The multi-step forecasts are the predicted future at time $t+1,t+2,\ldots,t+T$. We assume that the $T$-dimensional random vectors $\erbf(t) = (\er(t+1),\er(t+2),\ldots,\er(t+T))$ and $\elbf(t) = ( \el(t+1), \el(t+2), \ldots,\el(t+T))$ are jointly Gaussian with mean $\mu = (\mu_{\text{renew}}, \mu_{\text{load}})$ and covariance:
\begin{equation}
\Sigma = \begin{bmatrix} \Sigma_{\text{renew}} & \Sigma_{\text{renew,load}} \\
\Sigma^T_{\text{renew,load}}  & \Sigma_{\text{load}}
\end{bmatrix}.
\label{eq:forecast_errcov}
\end{equation}
A correlation is assumed between $\erbf$ and $\elbf$. This modeling choice is justified by the physical context of our study: solar energy is the RE source, and the solar irradiance is known to correlate with the cooling demand (AC load). 

When the forecast horizon $T$ is large (\eg, $T=672$), the sample covariance matrix $C$ becomes ill-conditioned. To mitigate this issue, we assume the errors $\erbf(t)$ and $\elbf(t)$ are wide-sense stationary. By partitioning the error vector into $D$ daily blocks (size $n_1 = 96$) with identical distributions, we impose a constant block-diagonal structure on $\Sigma_{\text{renew}}$ and $\Sigma_{\text{load}}$. This reduction in parameters transforms the estimation into a semidefinite program \cite{BoV:04}.
\begin{equation}
\begin{array}{ll}
\minimize & (1/2)\Vert \Sigma - C \Vert_F^2 \\
\text{subject to} & \Sigma \;\;\text{in \eqref{eq:forecast_errcov}} ,\;\;\Sigma \succ \epsilon I, \\
& \text{constant block diagonals in}\; \Sigma_{\text{renew}}, \Sigma_{\text{load}}
\end{array}
\label{eq:est_cove}
\end{equation}
with variable $\Sigma \in \reals^{2T\times 2T}$ and is symmetric.
The mean $\mu$ can be estimated from the sample mean. Once $(\hat{\mu}, \hat{\Sigma})$ are obtained, $(\erbf, \elbf)$ are jointly realized from the estimated Gaussian distribution. As $T$ can be large, \eg, in this study $2T = 1344$, solving~\eqref{eq:est_cove} can be numerically challenging. We apply the alternating direction method of multipliers (ADMM) \cite{BoydADMM} where the main update steps involve projecting a matrix to have constant block diagonals and a projection onto a positive definite cone; both of which can be efficiently computed. See the mathematical details in the Supplementary material.

\section{Results}

\subsection{Data description}
The data used in this project consists of two main components:
electrical load consumption and solar power generation from a single building, where it is capable of operating as a microgrid. The load data include electrical appliances and laboratory equipment, with a peak demand approximately 35 kWp. The generation data comprise solar irradiance, to develop forecasting models, and power output with a capacity of 15 kWp. The microgrid integrates a 100 kWh BESS with a 20–80\% SoC range (60\% usable capacity). EMS experiments were conducted using data from June 2024 to May 2025, aligning with the forecasting model's evaluation period. System parameters and electricity tariff are given in the Supplementary.

\subsection{Operation planning for RE day selection}
\label{sec:deterministic_planning_results}
In each batch of seven consecutive days, the deterministic planning results were produced by varying the daily CF compliance level from one to seven days, where only the mean forecasts of load and solar power from the best-performing model, LightGBM, were considered.

\begin{figure}
\centering
\begin{subfigure}{0.6\linewidth}
\centering
\includegraphics[width=0.95\linewidth]{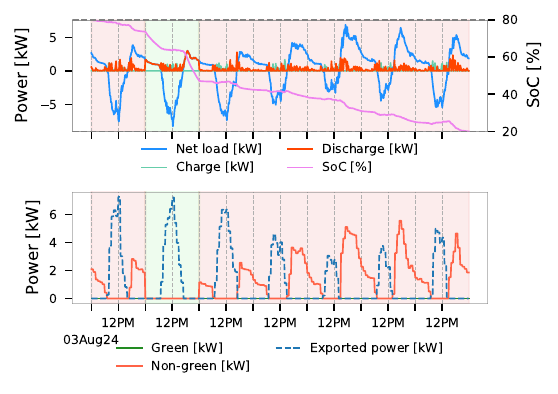}
\caption{One CF day (cost = 50 THB).}
\label{fig:plan7day_ts_1REdays}
\end{subfigure}

\begin{subfigure}{0.6\linewidth}
\centering
\includegraphics[width=\linewidth]{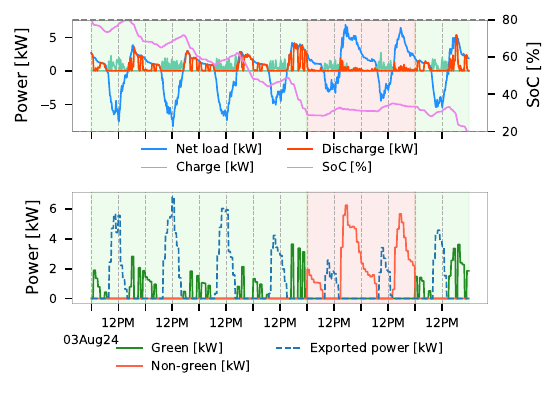}
\caption{Five CF days (cost = 77 THB).}
\label{fig:plan7day_ts_5REdays}
\end{subfigure}

\begin{subfigure}{0.6\linewidth}
\centering
\includegraphics[width=\linewidth]{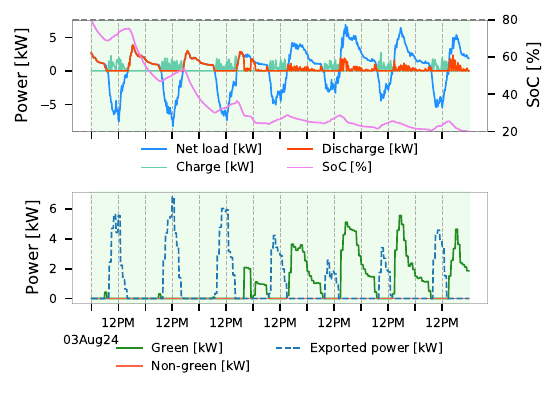}
\caption{Seven CF days (cost = 108 THB).}
\label{fig:plan7day_ts_7REdays}
\end{subfigure}
\caption{Optimal battery and purchased power planning to achieve the targeted number of CF days.}
\end{figure}

\Cref{fig:plan7day_ts_1REdays}-\Cref{fig:plan7day_ts_7REdays} 
are the results from the deterministic formulation \eqref{eq:cfeopt} with one green seller. For one CF day in \Cref{fig:plan7day_ts_1REdays}, the optimization selects the 2nd day, which has the smallest (negative) net load, as the CF day. On this day, the system relies solely on battery power and purchased green energy to meet the load demand. As the target increases to five CF days (\Cref{fig:plan7day_ts_5REdays}), the optimizer strategically designates days 5 and 6 as CB days. These days coincide with higher net load, where sourcing non-green energy is most cost-effective. Notably, the BESS discharges more actively on CF days (days 1–4 and 7) to satisfy the strict green-sourcing constraints. Finally, under a 7-day CF target, the system is forced to eliminate non-green energy reliance, instead maximizing battery utilization and minimizing green energy procurement to meet demand.

\begin{figure}
\centering
\includegraphics[width=0.7\linewidth]{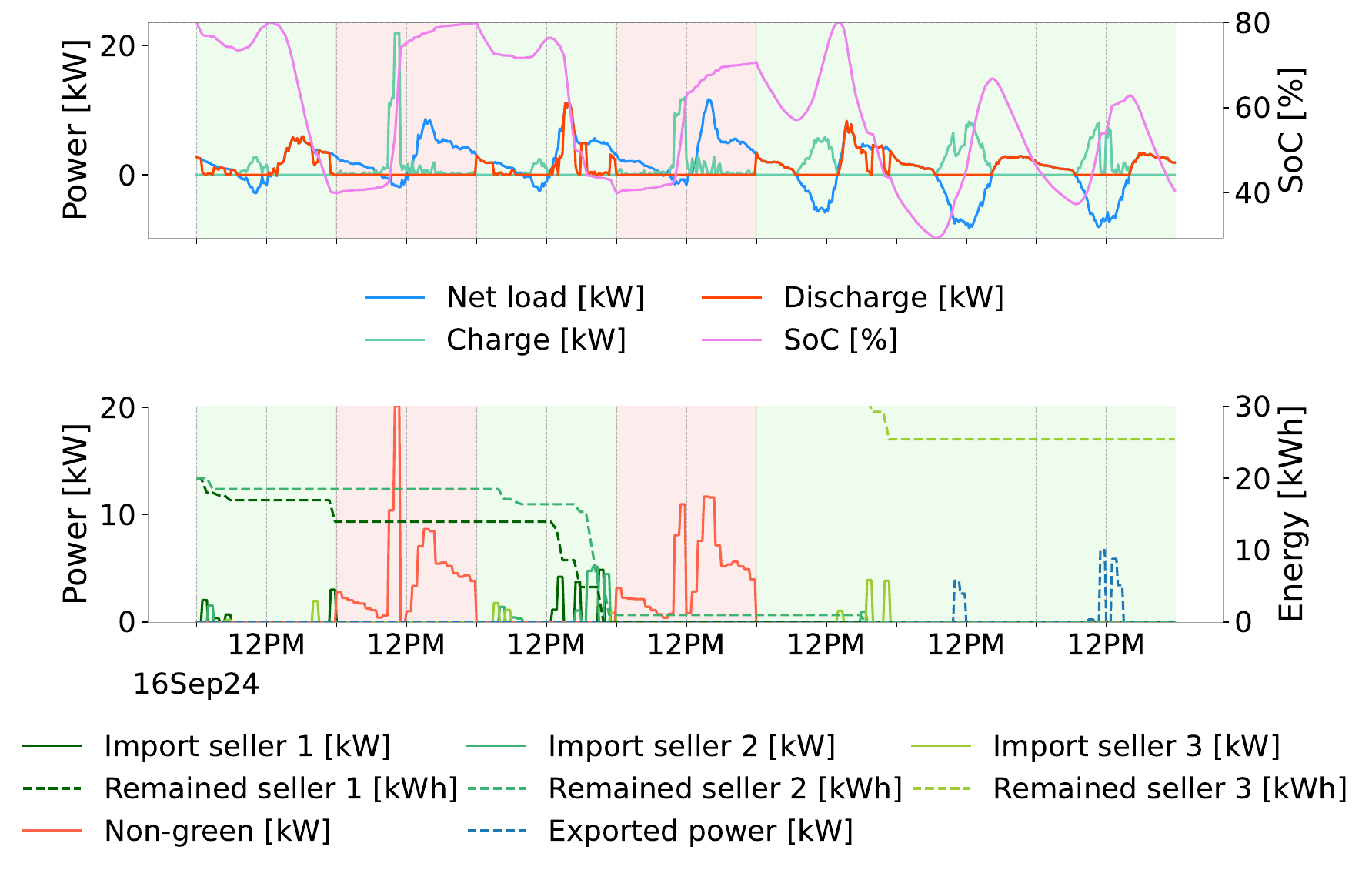}
\caption{Purchased power from different sellers.}
\label{fig:plan7day_purchase_power_multisource}
\end{figure}

Consider the multi-source case following \eqref{eq:greenbudget} with $\nsource = 3$. The green energy sellers are prioritized by unit price: Seller 1 is the least expensive, followed by Sellers 2 and 3. Regarding supply, Sellers 1 and 2 share identical initial quotas, while Seller 3 offers the highest capacity. \Cref{fig:plan7day_purchase_power_multisource} illustrates the multi-source procurement strategy for five CF days. The optimizer follows a merit-order dispatch: the lowest-priced Seller 1 is fully utilized first, followed by Seller 2 as its quota is reached. Seller 3, the most expensive, is dispatched only to meet remaining green energy demand. This hierarchy confirms a cost-optimal allocation subject to individual price and capacity constraints.

\subsection{Rolling Flexible CFE EMS}
To compare the performance of proposed formulations, both deterministic and stochastic EMS were implemented with a daily rolling approach. The battery charging/discharging and the purchasing/exporting of green and non-green energy were implemented according to the first-day solution. By imposing \eqref{eq:terminal_soc}, the initial SoC of the battery for the next planning day was not less than 40\% of the capacity. The EMS were run for one year period from June 2024 to May 2025 with the following settings:
\begin{itemize}
\item Ideal EMS: This case uses actual $\pr$ and $\pl$ measurements, representing perfect foresight. The resulting minimized net cost serves as the theoretical lower bound for all other forecast-based EMS strategies.
\item Deterministic EMS: The point forecasts of $\pr$ and $\pl$ from the best model were used in the planning.
\item Stochastic EMS: The scenarios of $\pr$ and $\pl$ were generated using the method in \eqref{eq:scenario_gen} with $\nscenario = 20$.
\end{itemize}

\textbf{Real-time adjustment.} To correct any mismatch between the planned and actual power balance after the uncertainty is revealed, all day-ahead commitments are treated as first-stage decisions.  Consequently, real-time adjustments are made only through the second-stage (real-time) variables: $\pgrti ,\pngrt ,\psrt $. Moreover, since the strict CF day requirements must be satisfied as planned, additional green power can be purchased only during the hours that were previously committed to green energy in the day-ahead stage. Similarly, non-green power can only be adjusted during the hours committed for non-green energy. The excess power, if any, can be sold back to the grid at the real-time selling price. All of the real-time prices are set at unfavorable rates compared to the day-ahead prices to discourage real-time transactions.

\begin{table}
    \centering
    \caption{Energy summary for CF targets from June 2024 to May 2025}
    \label{tab:rolling_energy_summary_all}
\begin{tabular}{lrrrrrrrrr}
\toprule
EMS & \multicolumn{2}{c}{Green [kWh]} & \multicolumn{2}{c}{Non-Green [kWh]} & \multicolumn{2}{c}{Grid Export [kWh]} & \multicolumn{2}{c}{Battery [kWh]} \\
 & DA & RT & DA & RT & DA & RT & Charge & Discharge \\
\midrule
Ideal CF71 & 7,948 & 0 & 8,405 & 0 & 9,393 & 0 & 4,863 & 4,090 \\
Deterministic CF71 & 7,849 & 8,120 & 5,061 & 2,842 & 11,182 & 6,060 & 2,782 & 2,339 \\
Stochastic CF71 & 6,257 & 6,430 & 7,036 & 2,209 & 2,190 & 13,508 & 338 & 292 \\
\midrule[0.4pt]
Ideal CF85 & 11,186 & 0 & 5,006 & 0 & 9,219 & 0 & 4,942 & 4,157 \\
Deterministic CF85 & 10,123 & 9,661 & 2,734 & 1,301 & 11,117 & 6,060 & 2,788 & 2,334 \\
Stochastic CF85 & 8,828 & 8,366 & 3,296 & 915 & 2,058 & 13,101 & 414 & 356 \\
\midrule[0.4pt]
Ideal CF100 & 15,683 & 0 & 0 & 0 & 8,658 & 0 & 5,262 & 4,425 \\
Deterministic CF100 & 12,655 & 10,963 & 0 & 0 & 10,906 & 6,060 & 2,930 & 2,466 \\
Stochastic CF100 & 11,216 & 9,803 & 0 & 0 & 1,990 & 12,775 & 456 & 391 \\
\bottomrule
\end{tabular}
\end{table}
  
\Cref{tab:rolling_energy_summary_all} summarizes the results for CF targets of five, six, and seven days, corresponding to CF71, CF85, and CF100 (24/7 CFE), respectively. The results show that the ideal EMS followed the planning outcomes exactly, with no need for imported green or non-green energy and no selling in real time. This occurred because perfect forecasts were available and real-time energy transactions were more expensive.

As the targeted number of CF days increased, the total energy imported from green sources in all EMS formulations also increased, while the imported non-green energy decreased. In addition, the required energy in real-time transactions of the deterministic EMS was the highest among all EMS formulations, indicating that relying on a single point forecast can lead to significant deviations from the planned operation.

When real-time adjustments are separated by energy type, the stochastic EMS consistently reduced both green and non-green real-time energy adjustments. Relative to the deterministic results, green RT energy decreased by $20.8\%$, $13.4\%$, and $10.6\%$ for CF71, CF85, and CF100, respectively. Non-green RT energy was also reduced by $22.3\%$ and $29.7\%$ for CF71 and CF85, while no non-green RT energy was required for CF100 in either EMS formulation.

The first-stage decisions were chosen to remain feasible across all scenarios, meaning that the solutions commit to actions that can withstand fluctuations in load and solar generation. This naturally results in a more conservative use of the battery in the day-ahead stage, since aggressive cycling cannot be guaranteed to remain feasible under every scenario.

\begin{figure}
    \centering
    \includegraphics[width=0.7\linewidth]{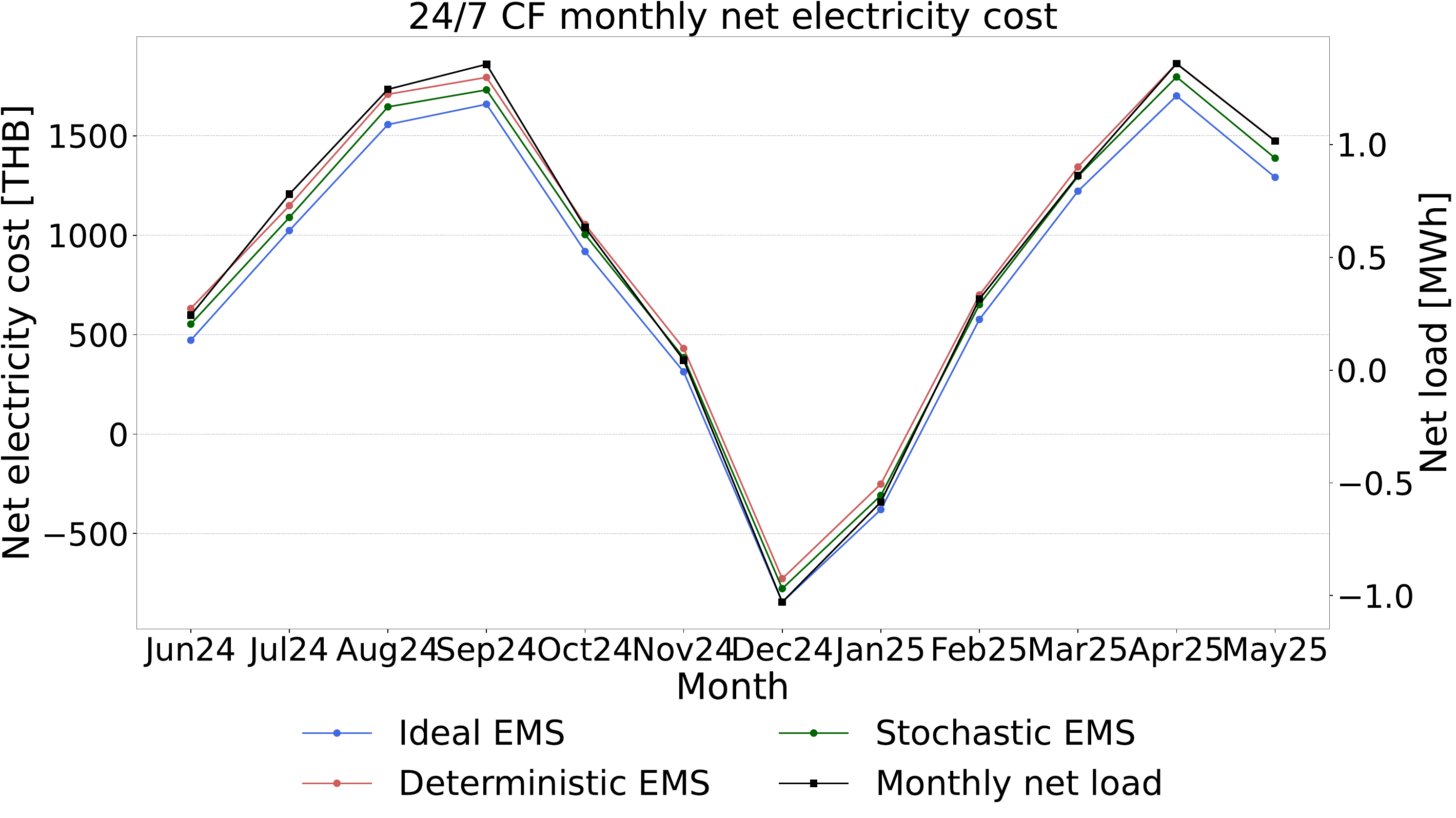}
    \caption{Monthly net cost comparison among different EMS formulations under 24/7 CF operation.}
    \label{fig:rolling_yearly_netcost.pdf}
    \end{figure}

For the 24/7 CFE target, \Cref{fig:rolling_yearly_netcost.pdf} provides monthly net electricity costs for all EMS formulations.
The cost pattern reflects the seasonal variations of both solar generation and student building load consumption. The stochastic EMS consistently achieves the closest net costs across all months to the ideal EMS, as compared to the deterministic EMS.

\begin{figure}
    \centering
    \includegraphics[width=0.7\linewidth]{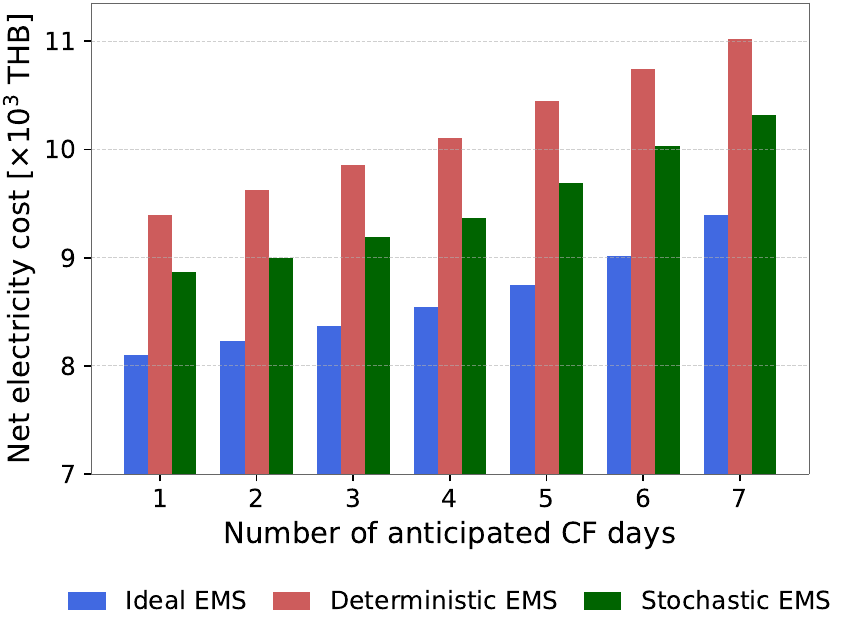}
    \caption{Yearly net cost comparison of rolling Flexible CFE EMS with different targeted CF days.}
    \label{fig:yearly_cost_bars}
    \end{figure}

\Cref{fig:yearly_cost_bars} summarizes the annual net electricity costs across CF day targets. The stochastic EMS consistently achieves lower net costs than the deterministic approach across all scenarios. While costs naturally increase as CF requirements become more stringent, the stochastic EMS effectively reduces these expenses, providing percentage reductions of  7.2\%, 6.6\%, and 6.4\% for the CF71, CF85, and CF100 targets, respectively, compared to the deterministic EMS.

Considering the obtained CF status over the full test period, it is important to note that even though the EMS formulation optimizes over a 7-day planning horizon each time, only the \emph{first-day} results are implemented before re-planning actions from the next day. This daily-rolling is constantly revised based on the most recent load and renewable forecasts become available. Consequently, the achieved \emph{daily CF percentages} in \Cref{tab:rolling_re_percentage} reflect only the implemented first-day decisions rather than the full multi-day horizon. Under this rolling framework, days designated for CF operation during planning may lose their CF status if subsequent re-planning shifts the optimal schedule.

\begin{table}
    \centering
    \caption{Achieved CF daily percentages where \textbf{boldface} values indicate performance exceeding the target.}
    \label{tab:rolling_re_percentage}
\begin{tabular}{lccc}
\toprule
Target CF & Ideal EMS & Deterministic EMS & Stochastic EMS \\
\midrule
CF14 & 12.6 & \textbf{14.5} & 12.3 \\
CF29 & \textbf{29.1} & \textbf{30.4} & 24.6 \\
CF43 & 41.9 & \textbf{45.8} & 38.5 \\
CF57 & 56.1 & \textbf{58.4} & 50.8 \\
CF71 & 69.0 & \textbf{74.6} & 68.7 \\
CF86 & 82.4 & \textbf{88.0} & \textbf{86.9} \\
CF100 & \textbf{100.0} & \textbf{100.0} & \textbf{100.0} \\
\bottomrule
\end{tabular}
\end{table}

The achieved CF percentages increase monotonically with anticipated CF level for all EMS formulations, indicating consistent alignment between planning objectives and operational outcomes, while some  deviations can still occur, as the status is determined from the executed first-day actions under re-planning rather than 
fixed multi-day commitments. The stochastic EMS exhibits comparatively lower achieved CF percentages for intermediate targets, which is aligned with its conservative decision that prioritizes feasibility across multiple scenarios. These deviations reflect operational limitations, where renewable commitments must be 
continuously revised as updated forecasts become available.

\begin{figure}
\centering
\includegraphics[width=0.7\linewidth]{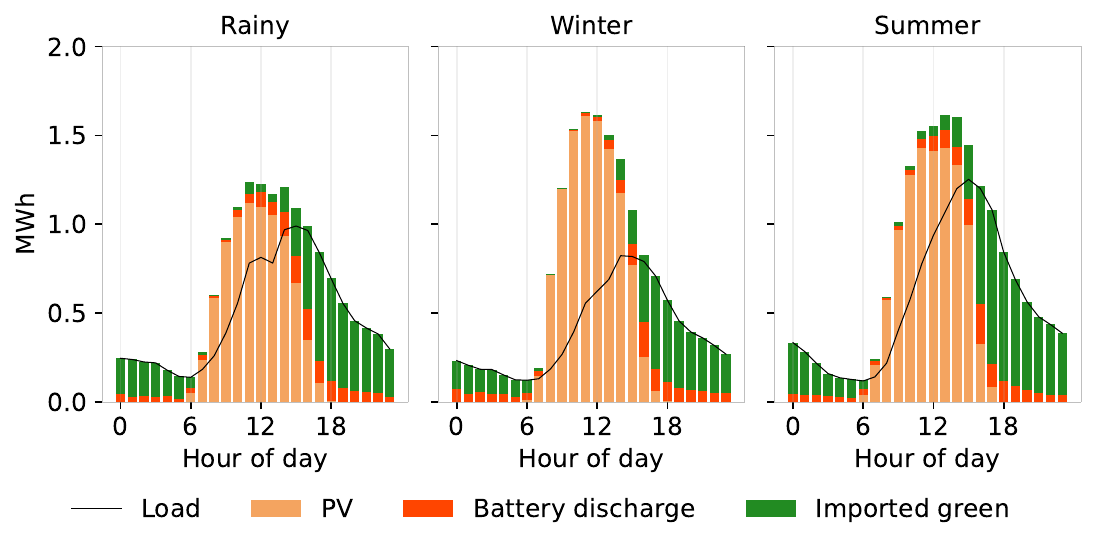}
\caption{Seasonal supply stacks and load for 24/7 operation.}
\label{fig:seasonal_supply}
\end{figure}

\Cref{fig:seasonal_supply} summarizes the seasonal supply stacks and corresponding load consumption profiles over the one-year test period under the ideal EMS of 24/7 CF operation, illustrating the hourly contributions of solar generation, battery discharge, and green power purchases used to meet load demand (office building). Clear seasonal patterns are observed on both the demand and supply sides, with higher solar generation in the summer and winter seasons compared to the rainy season. However, summer load demand is higher due to increased air-conditioning usage, resulting in a greater requirement for green energy supply to meet the 24/7 CF target. Seasonal variation has a significant impact on the energy required to maintain the 24/7 CF operation.

\section{Results: No excess energy export}
This section presents the optimization and experimental results for the Flexible CFE formulation under \textbf{no-sell} policy. These findings are presented in parallel with the \textbf{sell-back} policy results discussed in the main paper. Under a no-export policy, key results will show that battery utilization increases 2–35 times and electricity costs rise 6–7 times compared to the sell-back case. Furthermore, the achieved CFE percentages in the one-year rolling MPC frequently exceed target levels.

Considering the case where excess energy cannot be sold back to the grid, all formulations were modified by setting the sell rates to zero and excluding the relevant selling variables and terms in the objective functions and constraints. For the period that there exists excess energy after meeting the load demand and battery cannot be charged further, the excess energy is curtailed. The formulation changes include the following. The objective function of deterministic EMS becomes just the cost of importing energy:
\begin{equation}
\text{Cost} = \Delta t \sum_{t=1}^T  \left (  \bng(t) \png(t) +  \sum_{i=1}^{\nsource} \bg(t) \pg(t) \right ).
 \label{eq:netcost_nosell}
\end{equation}
For stochastic EMS, the variables $P_\text{sell,da}, P_\text{sell,rt}$ are removed from the day-ahead/real-time costs but a curtailment variable $P_\text{curt}(t,\xi)$ is added to the power balance constraint to account for any excess energy that cannot be exported. The modified objective functions are as follows.
\begin{align*}
f_{\text{da}}(x_{\text{1st}}) &= \Delta t \sum_{t=1}^T \left [ \bngda(t) \pngda(t) +  \sum_{i=1}^{\nsource} \bgda(t) \pgdai(t) \right ]. \\
f_{\text{rt}}(x_{\text{2nd}}(\xi)) &= \Delta t\sum_{t=1}^T  \bngrt(t) \pngrt(t,\xi) 
+ \Delta t \sum_{t=1}^T \sum_{i=1}^{\nsource} \bgrt(t) \pgrti (t,\xi) .
\end{align*}

For $t=1,\ldots,T$ and $\forall \xi \in \mathcal{S}$, 
\begin{multline}
\pc(t)  + \pl(t,\xi) + P_{\text{curt}}(t,\xi) = \sum_{i=1}^{\nsource} [ \pgdai(t) + \pgrti(t,\xi)] \\ 
+ \pngda(t) + \pngrt(t,\xi)   + \pdc(t) + \pr(t,\xi).
\label{eq:powerbalance_stoch_nosell}
\end{multline}

The next section reports the experimental results when there is no energy exported to the grid. 

\subsection{Deterministic EMS}

\begin{figure}
\centering
\begin{subfigure}{0.6\linewidth}
\centering
\includegraphics[width=0.95\linewidth]{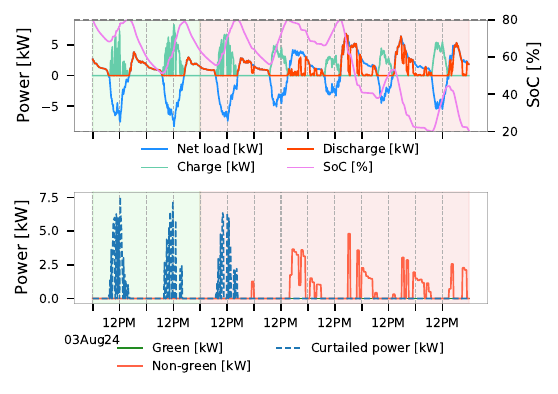}
\caption{One CF day (cost = 280 THB).}
\label{fig:plan7day_ts_1REdays_nosell}
\end{subfigure}

\begin{subfigure}{0.6\linewidth}
\centering
\includegraphics[width=\linewidth]{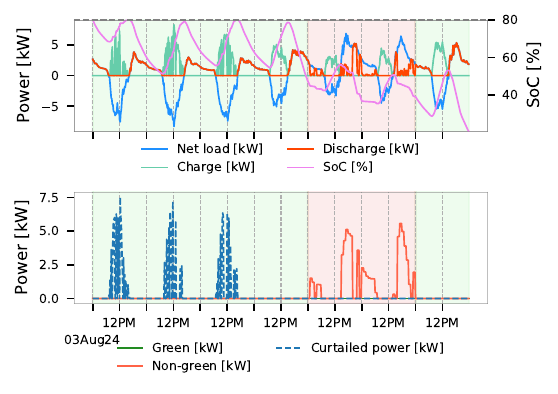}
\caption{Five CF days (cost = 280 THB).}
\label{fig:plan7day_ts_5REdays_nosell}
\end{subfigure}

\begin{subfigure}{0.6\linewidth}
\centering
\includegraphics[width=\linewidth]{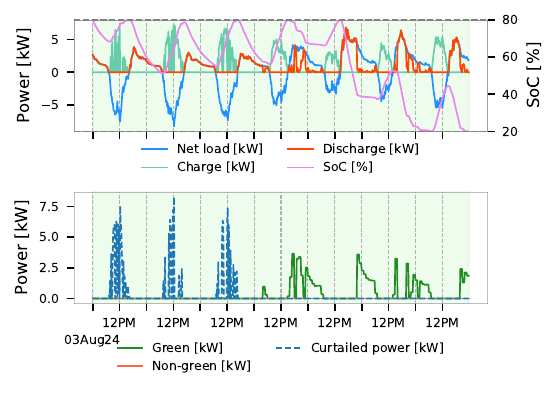}
\caption{Seven CF days (cost = 320 THB).}
\label{fig:plan7day_ts_7REdays_nosell}
\end{subfigure}
\caption{Optimal battery and purchased power planning to achieve the targeted number of CF days \textbf{without excess energy export}.}
\end{figure}

\Cref{fig:plan7day_ts_1REdays_nosell}-\Cref{fig:plan7day_ts_7REdays_nosell} 
are the results when varying targeted CF days to 1, 5 and 7 using a single green energy provider. Under the 1-day CF target, the system achieves two CF-compliant days (day 1 and day 2). For the 5-day CF target, the optimization selects day 1-4 and 7. These specific periods exhibit the significant negative net load, prompting the battery to be fully utilized and charged to its maximum SoC. This behavior represents a distinct shift in strategy compared to scenarios allowing energy export. Specifically, under a \textbf{sell-back} policy, the optimal operation for these initial days would prioritize exporting surplus power to the grid to maximize revenue; however, the \textbf{no-sell} constraint reorients the system toward local energy storage and CF compliance.

\begin{figure}
\centering
\includegraphics[width=0.7\linewidth]{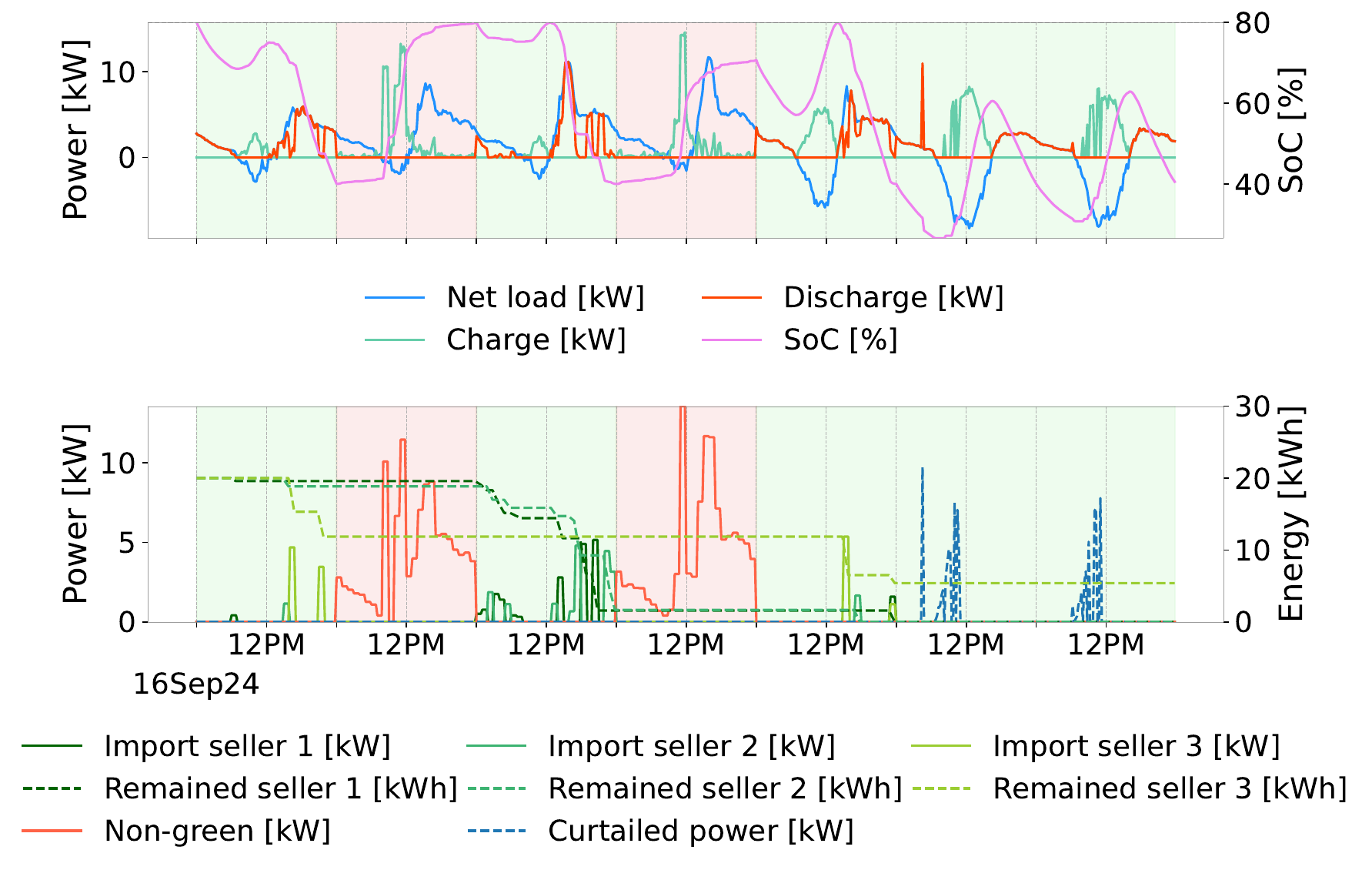}
\caption{Purchased power from different sellers \textbf{without excess energy export}.}
\label{fig:plan7day_purchase_power_multisource_nosell}
\end{figure}

\Cref{fig:plan7day_purchase_power_multisource_nosell} illustrates the multi-source procurement strategy for five CF days without excess energy sold back. The merit-order dispatch is still observed, where the lowest-priced Seller 1 is fully utilized first, followed by Seller 2 as its quota is reached. Seller 3, the most expensive, is dispatched only to meet remaining green energy demand.

\subsection{Rolling Flexible CFE EMS}
The same as the previous setting of the selling-back case, the rolling EMS were implemented with the daily re-planning framework.

\Cref{tab:rolling_energy_summary_all_nosell} summarizes the results for CF targets of five, six, and seven days, corresponding to CF71, CF85, and CF100 (24/7 CFE). Under the no-sell policy, all three EMS formulations significantly reduce grid imports and rely more heavily on battery operation to balance the system, as excess energy cannot be exported and must be either stored or curtailed. This naturally leads the operation to \textbf{prioritize battery usage} across all CF targets. Among the three formulations, the stochastic EMS commits the least DA non-green import and utilizes more on battery powers.

Across CF71 and CF85 targets, the stochastic EMS reduces DA nongreen imports by 73.8\% and 63.2\%, respectively, compared with the deterministic EMS. This reduction in day-ahead imports is partially offset by higher real-time procurement as uncertainty is revealed, while increased battery utilization is already scheduled in the first stage under the stochastic formulation, resulting in approximately 19–20\% higher battery discharge compared with the deterministic strategy.

\begin{figure}
    \centering
    \includegraphics[width=0.6\linewidth]{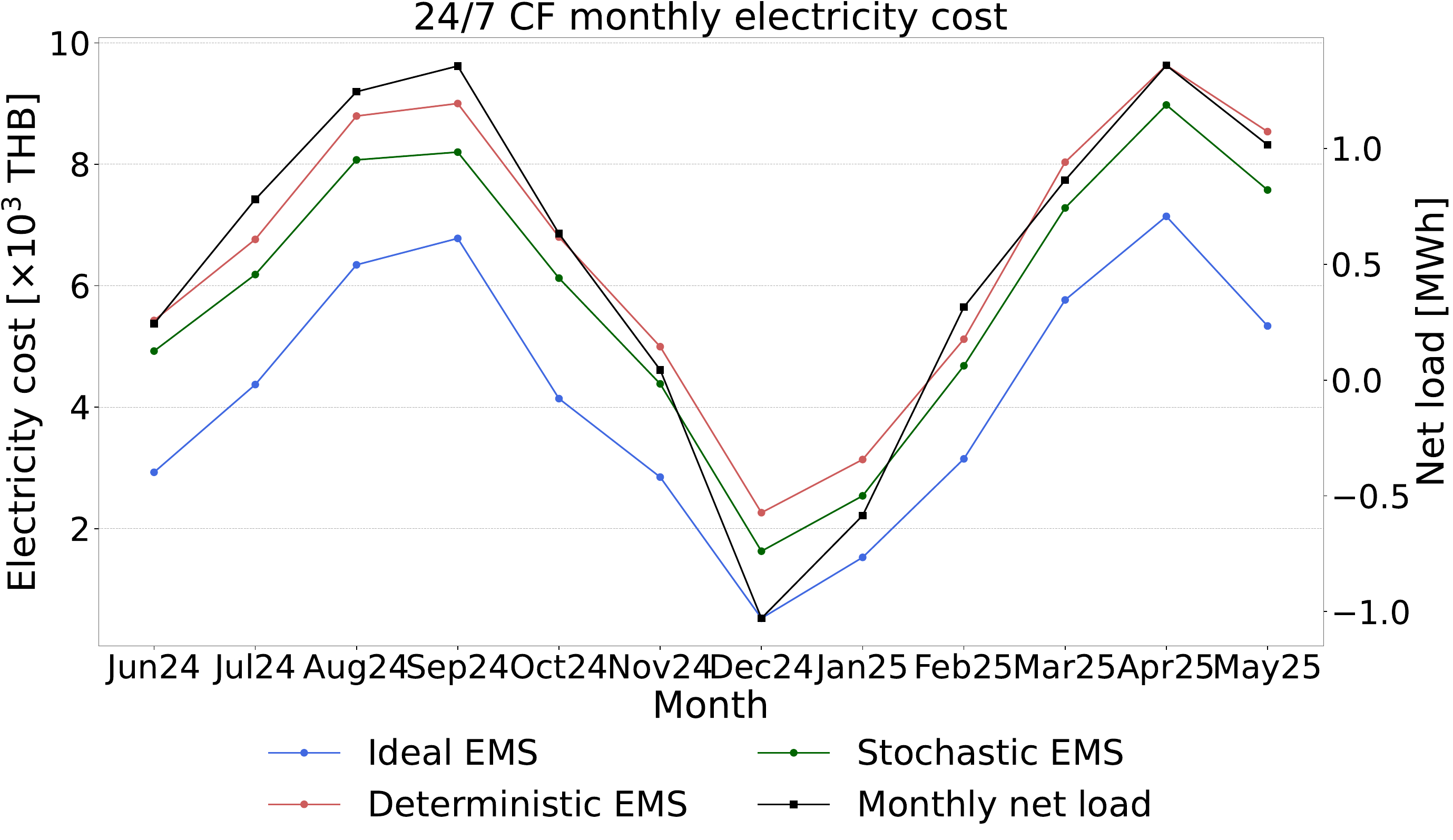}
    \caption{Monthly electricity cost comparison among different EMS formulations under 24/7 CF operation \textbf{without excess energy export}.}
    \label{fig:rolling_yearly_netcost_nosell}
    \end{figure}
    
Monthly electricity costs under the no-sell policy are also compared in \Cref{fig:rolling_yearly_netcost_nosell} for the 24/7 CF target. Overall, all EMS formulations show substantially higher electricity costs compared with the selling-back case, as excess energy can no longer be sold to the grid to offset costs. 

\begin{figure}
    \centering
    \includegraphics[width=0.6\linewidth]{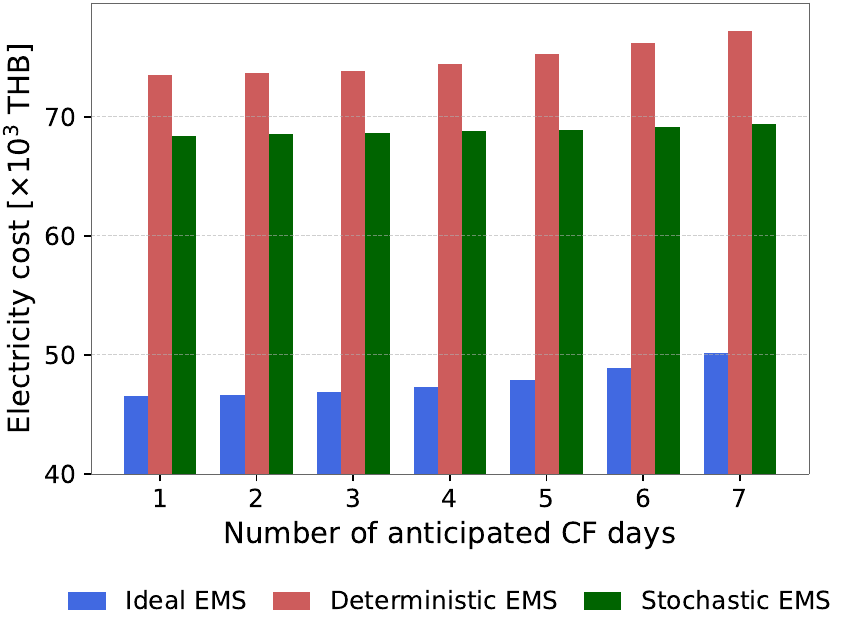}
    \caption{Yearly net cost comparison of rolling Flexible CFE EMS with different targeted CF days \textbf{without excess energy export}.}
    \label{fig:yearly_cost_bars_nosell}
    \end{figure}

\Cref{fig:yearly_cost_bars_nosell} shows annual electricity costs across various CF targets. Compared to the deterministic EMS, the stochastic approach reduces costs by 6.9–10.1\%, with greater savings as targets become more stringent. Without sell-back, however, costs increase 6-7 times.

\begin{table}
    \centering
    \caption{Energy summary for CF targets of \textbf{no-sell} policy from June 2024 to May 2025}
    \label{tab:rolling_energy_summary_all_nosell}
\begin{tabular}{lrrrrrrrr}
\toprule
EMS & \multicolumn{2}{c}{Green [kWh]} & \multicolumn{2}{c}{Non-Green [kWh]} & Curtailed [kWh] & \multicolumn{2}{c}{Battery [kWh]} \\
 & DA & RT & DA & RT & RT & Charge & Discharge \\
\midrule
Ideal CF71 & 3,941 & 0 & 7,107 & 0 & 3,109 & 10,951 & 9,200 \\
Deterministic CF71 & 2,448 & 7,623 & 3,758 & 2,676 & 8,602 & 10,677 & 8,961 \\
Stochastic CF71 & 4 & 11,160 & 983 & 2,595 & 6,510 & 12,726 & 10,680 \\
\midrule[0.4pt]
Ideal CF85 & 6,926 & 0 & 4,118 & 0 & 3,109 & 10,928 & 9,181 \\
Deterministic CF85 & 3,907 & 9,039 & 2,316 & 1,258 & 8,620 & 10,649 & 8,935 \\
Stochastic CF85 & 54 & 12,144 & 852 & 1,696 & 6,524 & 12,663 & 10,627 \\
\midrule[0.4pt]
Ideal CF100 & 11,026 & 0 & 0 & 0 & 3,109 & 10,810 & 9,082 \\
Deterministic CF100 & 6,158 & 10,361 & 0 & 0 & 8,617 & 10,686 & 8,972 \\
Stochastic CF100 & 259 & 14,366 & 0 & 0 & 6,375 & 12,834 & 10,771 \\
\bottomrule
\end{tabular}
\end{table}

\Cref{tab:rolling_re_daily_percentage_nosell} summarizes the achieved CF daily percentages. Without relying on excess energy export for maximizing profit, the achieved CF percentages under the no-sell policy are \textbf{consistently higher} than those obtained in the selling-back case and closely approach the targeted CF levels across all EMS formulations. This favorable outcome indicates that the no-sell policy more effectively enforces CF compliance by constraining grid transactions and prioritizing battery energy storage utilization, leading to a more reliable realization of 24/7 CF operation.

\begin{table}
    \centering
    \caption{Achieved CF daily percentages of \textbf{no-sell} policy where \textbf{boldface} values indicate performance exceeding the target.}
    \label{tab:rolling_re_daily_percentage_nosell}
\begin{tabular}{lccc}
\toprule
Target CF & Ideal EMS & Deterministic EMS & Stochastic EMS \\
\midrule
CF14 & \textbf{29.1} & \textbf{37.4} & \textbf{77.1} \\
CF29 & \textbf{38.5} & \textbf{42.7} & \textbf{79.3} \\
CF43 & \textbf{46.6} & \textbf{51.1} & \textbf{80.7} \\
CF57 & \textbf{58.9} & \textbf{63.1} & \textbf{83.0} \\
CF71 & 70.7 & \textbf{77.4} & \textbf{86.6} \\
CF86 & 84.9 & \textbf{88.5} & \textbf{91.1} \\
CF100 & \textbf{100.0} & \textbf{100.0} & \textbf{100.0} \\
\bottomrule
\end{tabular}    
\end{table}

\begin{figure}
    \centering
    \includegraphics[width=0.7\linewidth]{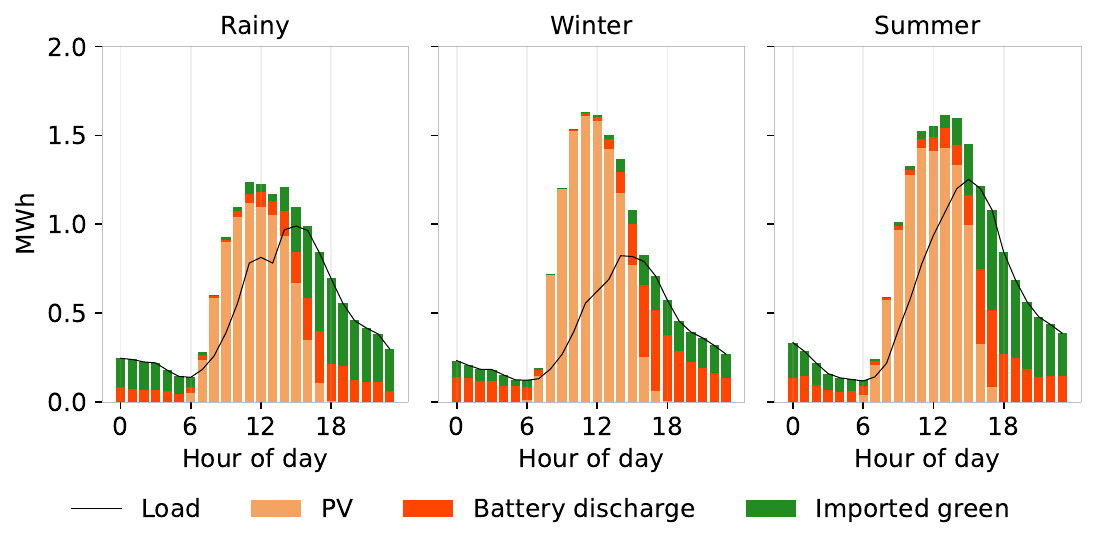}
    \caption{Seasonal supply stacks and load for 24/7 operation \textbf{without excess energy export}.}
    \label{fig:seasonal_supply_nosell}
\end{figure}

Another evidence of the impact of no-sell policy on more relying on battery operation is illustrated in \Cref{fig:seasonal_supply_nosell}, which summarizes the seasonal supply stacks and corresponding load consumption profiles over the one-year period under the ideal EMS of 24/7 CF operation without excess energy export. 
Compared with the sell-back policy, the battery discharge contribution is significantly increased to meet the load demand, while the purchased green energy is reduced. This observation confirms that the no-sell policy encourages more local energy storage utilization to achieve CF compliance.

\section{Conclusion} 
This paper illustrates how to plan and operate EMS to achieve CFE goals for the building microgrid with BESS and solar PV components.
The strategy involves selecting specific days for CF operation within a daily rolling planning framework, optimizing battery usage and energy procurement to minimize costs while adhering to CF requirements. The CF day selection strategy is devised by formulating an MILP with a group $\ell_\infty$-norm constraint on a limited number of CB (carbon-based) days, where this idea can be generalized to selecting any CF interval (hourly/15-minute) within a specified compliance. The results demonstrate that the stochastic flexible CFE formulation effectively reduces real-time energy adjustments and has lower net costs about 6.4-7.2\% compared to the deterministic approach across various CF targets. Finally, a practical limitation arises from the rolling EMS implementation, in which only first-day decisions are executed before re-planning, leading to potential deviations from the planned CF status. This limitation is inherent to operating under evolving information and is standard in rolling-horizon EMS frameworks.

\section*{Acknowledgment}
This research project is financially supported by the Ratchadaphiseksomphot Endowment Fund, Chulalongkorn University. During the preparation of this work the authors used Gemini 3 Flash in order to improve readability and language. After using this tool, the authors reviewed and edited the content as needed and took full responsibility for the content of the published article.

\bibliographystyle{alpha}
\bibliography{re100ems}
\addcontentsline{toc}{section}{References}

\appendix

\newpage
\section*{Supplementary Material}
\addcontentsline{toc}{section}{Supplement}

\section{EMS parameters and electricity tariff}
The battery system parameters, the buy and sell rates are specified in \cref{tab:sysparam} and \cref{tab:tariff}. 

\begin{table}[ht]
\centering
\caption{System and battery parameters.}
\begin{tabular}{lll}
\toprule
Parameter & Value & Unit \\
\midrule
Electrical load peak & 35 & kWp \\
PV capacity & 15 & kWp \\
BattCapacity & 100 & kWh \\
Charging efficiency $\eta_c$ & 0.95 & - \\
Discharging efficiency $\eta_d$ & 0.8835 & - \\
$\text{Charge rate}_{\max}$ & 50 & kW \\
$\text{Discharge rate}_{\max}$ & 50 & kW \\
$\soc_{\min} $ & 20 & \% \\
$\soc_{\max} $ & 80 & \% \\
Max charging ramp rate $P_\text{chg,ramp}$ & 20 & kW \\
Max discharging ramp rate $P_\text{dchg,ramp}$ & 20 & kW \\
\bottomrule
\end{tabular}
\label{tab:sysparam}
\end{table}

\begin{table}[ht]
\centering
\caption{Electricity tariff rates (THB). DA: Day-ahead rate. RT: Real-time rate.}
\begin{tabular}{l|rrrr}
\toprule
Rate & Non-green & Green no.1 & Green no.2 & Green no.3 \\
\midrule
DA Buy rate & 4.22 & 4.55 & 5.20 & 5.75 \\
RT Buy rate & 4.50 & 4.75 & 5.40 & 5.95 \\ 
\midrule
DA Sell rate & \multicolumn{4}{c}{3.90} \\
RT Sell rate & \multicolumn{4}{c}{3.80} \\
\bottomrule
\end{tabular}
\label{tab:tariff}
\end{table}
The non-green energy buy rate is always less than the green energy. The green procurement rates are sorted ascendingly from seller 1 through 3. The DA rates are systematically more favorable than RT rates; specifically, buy rates are lower and sell rates are higher in the DA market compared to the RT market. Last, the buy rates remain systematically higher than sell rates. For the single-seller experiment, Green no. 1 was utilized as the primary reference.

\section{Covariance estimation with equal block diagonals}
Let $\Sigma$ be the covariance of forecast error $(\erbf,\elbf) \in \reals^{T} \times \reals^{T}$ taking the form:
\begin{equation}
\Sigma = \begin{bmatrix} \Sigma_{\text{renew}} & \Sigma_{\text{renew,load}} \\
\Sigma^T_{\text{renew,load}}  & \Sigma_{\text{load}}
\end{bmatrix}.
\end{equation}
Let $C$ be the sample covariance that can be calculated from forecast errors obtained from the training set. Due to high-dimensional-low-sample setting, $C$ is often ill-conditioned. Moreover, we have some prior structure on $\Sigma$. The problem of estimating $\Sigma$ with equal block diagonals is proposed as
\begin{equation}
\begin{array}{ll}
\minimize & (1/2)\Vert \Sigma - C \Vert_F^2 \\
\text{subject to} &  \Sigma \succ \epsilon I, \\
& \text{constant block diagonals in}\; \Sigma_{\text{renew}}, \Sigma_{\text{load}}
\end{array}
\label{eq:est_cov_supplement}
\end{equation}
with variable $\Sigma \in \reals^{2T\times 2T}$ and is symmetric. This is a semidefinite program and it is numerically challenging to solve when $T$ is large. This section describes the details of implementing the ADMM algorithm to solve the problem.

Denote $\symm^n$ the set of symmetric matrices of size $n \times n$. Define the two sets:
\begin{gather*}
\mathcal{M}_1 = \left \{ X \in \symm^{2n}  \;|\; X = \begin{bmatrix} X_{11} & X_{12} \\ X_{12}^T & X_{22} \end{bmatrix} , \;\; 
 X_{11}, X_{22} \in \symm^n \;\; \text{have block diagonals of size $r < n$} \right \} \\
\mathcal{M}_2 = \left \{ X \in \symm^{2n} \;|\; X  \succeq \epsilon I \right \} \;\;\text{for $\epsilon > 0$}.
\end{gather*}
Define the two functions:
\begin{align}
f(X) &= (1/2) \Vert X - C \Vert_F^2, \;\;\text{with $\dom f(X) = \mathcal{M}_1$} \\
g(Z) &= I_{\mathcal{M}_2} (Z) \;\; \text{(indicator function)}.
\end{align}
The problem~\eqref{eq:est_cov_supplement} can be arranged into the ADMM format as 
\begin{equation}
\begin{array}{ll}
\minimize & f(X) + g(Z) \\
\mbox{subject to} & X - Z = 0.
\end{array}
\label{eq:est_cov_admm}
\end{equation}
with variables $X,Z \in \symm^{2n}$. The augmented Lagrangian (scaled form) is 
\[
L(X,Z,U) = f(X) + I_{\mathcal{M}_2} (Z) + \rho\Tr( U^T(X-Z)) + (\rho/2) \Vert X - Z \Vert_F^2,
\]
where $\rho > 0$ is the ADMM parameter. With the superscript $k$ as the iteration index, the ADMM update rules are as follows. \\[2ex]

{\color{teal}\hrule height 0.8pt}
\vspace{1mm}
\textbf{ADMM algorithm for problem~\eqref{eq:est_cov_supplement}} \\
\vspace{-2mm}
{\color{teal}\hrule height 0.8pt}

\begin{itemize}
\item Initialize $X^0 = C, Z^0 = C, U^0 = 0 $, and set $\rho > 0$.
\item $X$-update
\[
X^{k+1} = \argmin_{X \in \mathcal{M}_1} \; \frac{1}{2} \Vert X - C \Vert_F^2 + \frac{\rho}{2} \Vert X - Z^{k} +U^{k} \Vert_F^2
\]
By completing the square, the objective function can be rewritten as
\[
\left \Vert X - \frac{C+\rho(Z^k-U^k)}{\rho+1} \right \Vert_F^2 
\]
The $X$-update is then a problem of projection onto $\mathcal{M}_1$.
\[
X^{k+1} = \Pi_{\mathcal{M}_1} \left [  \frac{1}{\rho+1} ( C + \rho(Z^k-U^k)) \right ],
\]
which is computationally cheap because the constraint in $\mathcal{M}_1$ is simply linear. The projection has a closed-form solution where $\Pi_{\mathcal{M}_1}[W]$ is given by $X$ whose all entries are equal to $W$, except the block $X_{11}, X_{22}$ are obtained by taking the average of block diagonals in $W_{11}$ and $W_{22}$, respectively. 
\item $Z$-update 
\begin{align*}
Z^{k+1} &= \argmin_{Z \in \mathcal{M}_2} \;\; (\rho/2) \Vert X^{k+1} - Z + U^k \Vert_F^2  \\
&= \Pi_{Z \succeq \epsilon I} [ X^{k+1} + U^k ]
\end{align*}
The projection on $\mathcal{M}_2$ is obtained by performing eigenvalue decomposition of a symmetric matrix where the eigenvalues are clipped by the threshold $\epsilon$, \ie, Let $W = VDV^T$ where $D = \diag(d)$. Then, $\Pi_{\mathcal{M}_2}[W] = VD^{+}V^T$ where $D^{+} = \diag(\max(\epsilon,d))$. 
\item $U$-update: $U^{k+1} = U^k + X^k - Z^k$.
\end{itemize}

{\color{teal}\hrule height 0.8pt} 
\vspace{2mm}
Overall, the $X$- and $U$-update steps are just matrix additions while the most expensive part is the eigenvalue decomposition in the $Z$-update step. We repeat the update step until the primal and dual residuals are less than a specified tolerance.

\end{document}